\documentclass{jpp}

\usepackage{graphicx}
\usepackage{physics}

\usepackage[utf8]{inputenc}
\usepackage[T1]{fontenc}
\usepackage{amsmath}

\usepackage{xcolor}

\shorttitle{Dissipation at intermittent structures}
\shortauthor{T. H. N. Tsung, G. R. Werner, D. A. Uzdensky, M. C. Begelman}

\title{Dissipation and particle acceleration at intermittent structures with velocity and magnetic shear: Interaction of Kelvin-Helmholtz and Drift-Kink instabilities}

\author{Tsun Hin Navin Tsung\aff{1}$^,$\aff{3}
  \corresp{\email{tsunhinnavin.tsung@colorado.edu}},
  Gregory R. Werner\aff{1},
  Dmitri A. Uzdensky\aff{1}$^,$\aff{2},
 \and Mitchell C. Begelman\aff{3}$^,$\aff{4}}

\affiliation{\aff{1}Center for Integrated Plasma Studies, Physics Department, 390 UCB, University of Colorado, Boulder, CO 80309, USA
\aff{2}Rudolf Peierls Centre for Theoretical Physics, Clarendon Laboratory, University of Oxford, Parks Road, Oxford OX1 3PU, UK
\aff{3}JILA, University of Colorado and National Institute of Standards and Technology, 440 UCB, Boulder, CO 80309-0440, USA
\aff{4}Department of Astrophysical and Planetary Sciences, University of Colorado, 391 UCB, Boulder, CO 80309, USA}

\begin{document}

\maketitle

\begin{abstract}
We present two-dimensional (2D) particle-in-cell simulations of a magnetized, collisionless, relativistic pair plasma subjected to combined velocity and magnetic-field shear, a scenario typical at intermittent structures in plasma turbulence. We create conditions where only the Kelvin-Helmholtz (KH) and Drift-Kink (DK) instabilities can develop, while tearing modes are forbidden. The interaction of DKI and KHI generates qualitatively new structures, marked by a thickened shear layer with very weak electromagnetic field, modulated by KH vortices. Over a range of moderately strong velocity shears explored, the interaction of DKI and KHI results in a significant enhancement of dissipation over cases with only velocity shear or only magnetic shear. Moreover, we observe a new and efficient way of particle acceleration where particles are stochastically accelerated by the motional electric field exterior to the shear layer as they meander in an S-shaped pattern in and out of it. This process takes advantage of the bent geometry of the shear layer caused by the DK-KHI interaction and is responsible for most of the highest-energy particles produced in our simulations. These results further our understanding of dissipation and particle acceleration at intermittent structures, which are present in plasma turbulence across a wide range of astrophysical contexts such as in AGN jet sheaths, potentially relevant to limb-brightened emission, etc., and highlight the sensitivity of dissipation to multiple interacting instabilities, thus providing a strong motivation for further studies of their nonlinear interaction at the kinetic level.
\end{abstract}

\section{Introduction} \label{sec:introduction}

Intermittency, defined as the inherent spatiotemporal inhomogeneity of turbulence due to random fluctuations in the energy cascade as it proceeds from large to small scales \citep{Zhdankin_etal-2016}, is generally present in astrophysical plasma turbulence. It manifests as thin structures with enhanced current densities and/or vorticities that take up a small fraction of space but account for a substantial portion of the energy dissipated in turbulence \citep{Zhdankin_etal-2016}. In a magnetized plasma, these structures imply strong velocity and magnetic shears coinciding in space. These two shears drive various instabilities. Thus, velocity shear is prone to the Kelvin-Helmholtz (KH) instability, which has been studied extensively using linear theory, 
fluid simulations 
and kinetic simulations. 
On the other hand, magnetic shear is susceptible to tearing modes and magnetic reconnection 
creating plasmoid chains and accelerating particles efficiently \citep{Zenitani_Hoshino-2007,Guo_etal-2014,Werner_etal-2016,Sironi_etal-2025}. Magnetic shear is also prone to Drift-Kink (DK) instability \citep{Zhu_Winglee-1996,Daughton-1999,Pritchett_etal-1996,Zenitani_Hoshino-2007,Zenitani_Hoshino-2008,Barkov_Komissarov-2016, Werner_Uzdensky-2021} which, in a pair plasma with a weak guide magnetic field, grows faster during the linear stage than the tearing modes, but is generally not as efficient a particle accelerator as reconnection \citep{Zenitani_Hoshino-2007}. How the instabilities arising from velocity and magnetic shear interact with one another is an open question that has only now begun to be explored. A practical question is how such interaction affects energy dissipation and nonthermal particle acceleration (NTPA). With applications to the boundaries of relativistic jets in active galactic nuclei (AGN) in mind, \citet{Sironi_etal-2021} considered a 2D jet-wind model --- a `jet' medium made up of pair plasma and a `wind' medium made up of normal (electron-ion) plasma with a relativistic velocity shear between them, and a magnetic field that was helical in the jet and toroidal and significantly weaker in the wind. That study found that KH vortices can wrap the field lines over each other, creating current sheets that trigger reconnection and considerable dissipation, but the specific setup of that study precluded the authors from exploring the effects of~DKI. 

\begin{figure}
    \centering
    \includegraphics[width=0.8\textwidth]{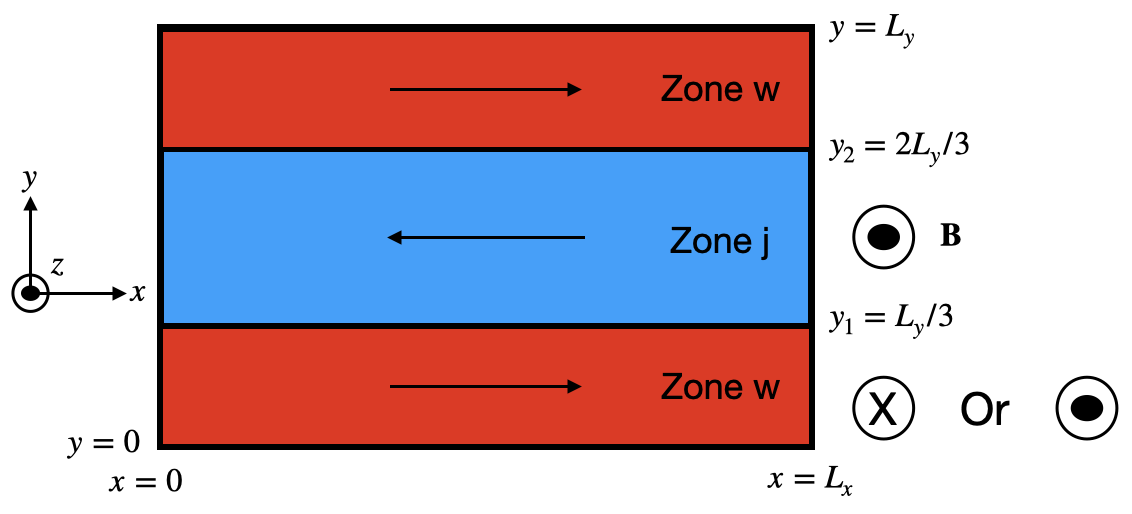}
    \caption{The double shear layer setup used in this study, consisting of two zones, Zone j and Zone w, in equal and opposite motion along the $x$-axis with each other. The magnetic field in Zone j points out of the page ($+z$ direction) while that in Zone w can point either into ($-z$) or out of the page. The plasma quantities $B, E$, etc. are connected smoothly between the two zones by a tanh profile.
    }
    \label{fig:PIC_setup}
\end{figure}

In this article we consider a similar computational setup (see Fig.~\ref{fig:PIC_setup}), but simplified in order to explore the joint effects of KH and DK modes, and their nonlinear interplay, on magnetic and bulk-kinetic dissipation. Here, for the first time, we incorporate the effects of velocity and magnetic shear in a 2D configuration where DKI coexists with~KHI, while tearing is forbidden. Using first-principles particle-in-cell (PIC) simulations, we show that the interaction between the KH and DK instabilities creates qualitatively new structures and very different amounts of dissipation compared to the case when only one of the instabilities is present. It also leads to a new and efficient means of nonthermal particle acceleration. Thus, dissipation is highly sensitive to the interplay of the two instabilities.

The structure of this paper is as follows: in \S\ref{sec:methods} we describe the simulation setup for this study, including the parameters explored and the diagnostics used. In \S\ref{sec:results} we present our results, focusing first on the morphology and dissipation of the system in~\S\ref{subsec:morp_diss} and then on particle acceleration in~\S\ref{subsec:ptcl_acc}. In \S\ref{sec:conclusion} we discuss the applicability of our results and draw our conclusions.


\section{Methods} \label{sec:methods}

We perform 2D relativistic electromagnetic PIC  simulations using the code {\sc Zeltron}~\citep{Cerutti_etal-2013}. We consider a relativistically warm ($\theta \equiv T/m_e c^2\sim1$) electron-positron pair plasma in a double-layer shear flow (see Fig.~\ref{fig:PIC_setup}). 
Our setup consists of two regions, `Zone j' (blue) and `Zone w' (red), with two thin interface layers between them. In the following, $x$ is the direction of the fluid flow while $y$ is the gradient direction. We impose an initial out-of-plane magnetic field that is perpendicular to the flow and the gradient direction, i.e. $\vb{B} = B(y)\vu{z}$. We simulate the $x,y$-plane keeping all three components for vector quantities. Since the $z$-direction is not simulated, no tearing modes can be excited (a full 3D study is left to future work). This allows us to focus on the KH and DK modes and their interplay. 
We run the simulations in the frame where both zones are moving at the same speed but in opposite ($\pm \vu{x}$) directions. 

The setup is characterized by the following initial dimensionless parameters: Zone j's bulk-flow 4-velocity $u_j = \Gamma_j\beta_j$, normalized temperature $\theta_j= T_j/m_e c^2$, magnetization $\sigma_j = B_j^2/4\pi w_j$ (where $w_j$ is Zone j's relativistic enthalpy density), and the ratio of Zone w's to Zone j's magnetic field~$B_w/B_j$. Zone j's magnetization $\sigma_j$ and normalized temperature $\theta_j$ are set to 1 initially in all the cases we consider, while $u_j$ is varied from run to run. Zone w's rest-frame density, temperature, and bulk speed are initially set equal to those in Zone~j (with bulk velocity reversing direction). We connect Zone j's and Zone w's flow velocities and magnetic fields smoothly by a tanh function of $y$ with a transition half-width~$\Delta$. 
Combining Maxwell's equations and the ideal gas law with the assumptions of pressure balance across the shear layer and the ideal-MHD condition that the initial electric field is purely motional, $\vb{E} = -\vb{v}\cross\vb{B}/c$, the $y$-profiles of the electric field, velocity, density, and temperature of the electrons and positrons can be determined. The initial velocities of the particles are then sampled from the local drifting Maxwell-J\"uttner (MJ) distribution. Details for this setup can be found in Appendix \ref{app:setup}. With this setup, we ran two control cases with only one type of shear present $(u_j, B_w/B_j) = (0.5,+1)$, and $(0, -1)$ (hereafter referred to as the VS and MS case, respectively), and a series of mixed cases with fixed magnetic shear ($B_w/B_j = -1$) and progressively varied~$u_j:\{0.01,0.03,0.05,0.07,0.1,0.2,0.3,0.4,0.5,0.6,0.7,0.8,0.9\}$. This allowed us to investigate systematically the effects of velocity shear on DKI and its co-evolution with KHI.

The simulation domain is a rectangular box with periodic boundaries spanning $L_x = 100 d_{e,j}$ in the $x$-direction and $L_y = 300 d_{e,j}$ in the $y$-direction; it is resolved fiducially by $1024\times 3072$ grid cells, which resolves Zone j's electron inertial length~$d_{e,j}$ by $\approx$10 cells. The initial half-thickness of each  shear layers, located at $y_{1,2} = \{L_y/3, 2L_y/3\} = \{100, 200\}\, d_{e,j}$, is set to $\Delta = 5 d_{e,j}$ as motivated by \citet{Serrano_etal-2024}, which finds that intermittent structures typically have a width of~$3 d_{e,j}$. For relativistically warm ($\theta\sim 1$), moderately magnetized ($\sigma\sim 1$) pair plasma, Zone j's electron Debye length~$\lambda_{\mathrm{D}e}$, average gyroradius~$\rho_{e,j}$, and  electron inertial length~$d_{e,j}$ are roughly equal, so all the kinetic scales are well resolved (exact definitions for~$d_{e,j},\lambda_{\mathrm{D},e},\rho_{e,j}$ are given in Appendix~\ref{app:microscales}, though we quote that $\rho_{e,j}\sim (\theta_j/\sigma_j)^{1/2}d_{e,j}$). We create 64 particles per cell (PPC) per species (totaling 128~PPC); this results in $d^2_{e,j}/\Delta x\Delta y\times \mathrm{PPC} \sim 6400$ particles per area spanned by Zone j's electron skin depth, sufficient for our purposes. We run our simulations for approximately $50 L_x/c$ to ensure the saturated stage is reached. For kinetic diagnostic purposes, we randomly select and track the trajectories of $10^5$ electrons and positrons throughout the simulation.

The main focus of this study is on the evolution of the total magnetic and bulk kinetic energies. The total magnetic energy is calculated as $E_B \equiv \int B^2/8\pi\,\mathrm{d}V$. To calculate the total bulk kinetic energy, we first perform a Lorentz transformation into the local zero-particle-flux (Eckart) frame at each cell, 
calculate the local pressure tensor there, and then subtract it from the total stress tensor to obtain the bulk-flow stress tensor. 
Integrating its trace over the box volume gives the total bulk kinetic energy $E_\mathrm{KE}$ (details of this procedure are given in Appendix \ref{app:decompose}). As a consistency check, the total energy in all our simulations is conserved to within~0.2\%.

\begin{figure}
    \centering
    \includegraphics[width=0.7\textwidth]{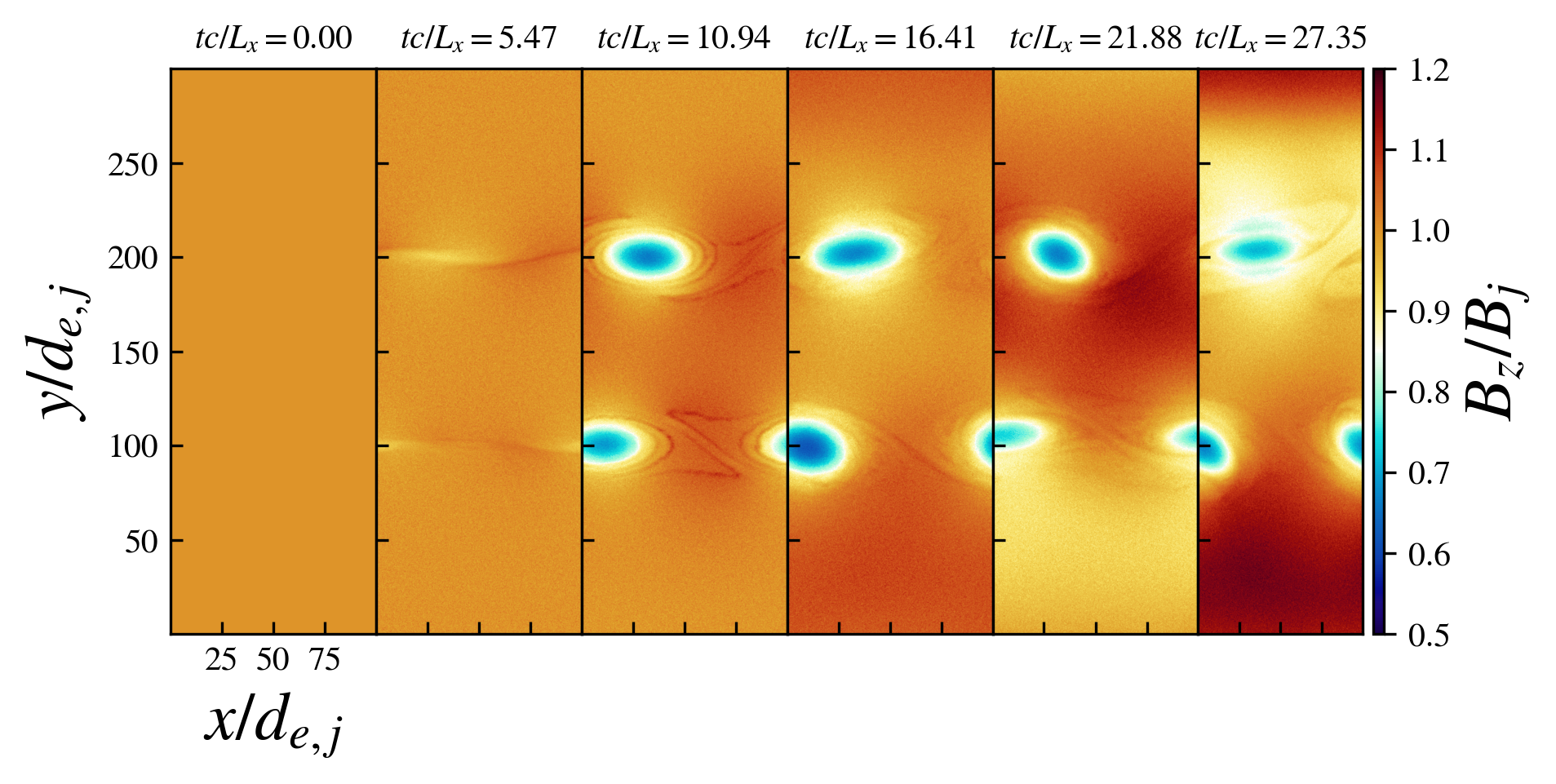} \\
    \includegraphics[width=0.7\textwidth]{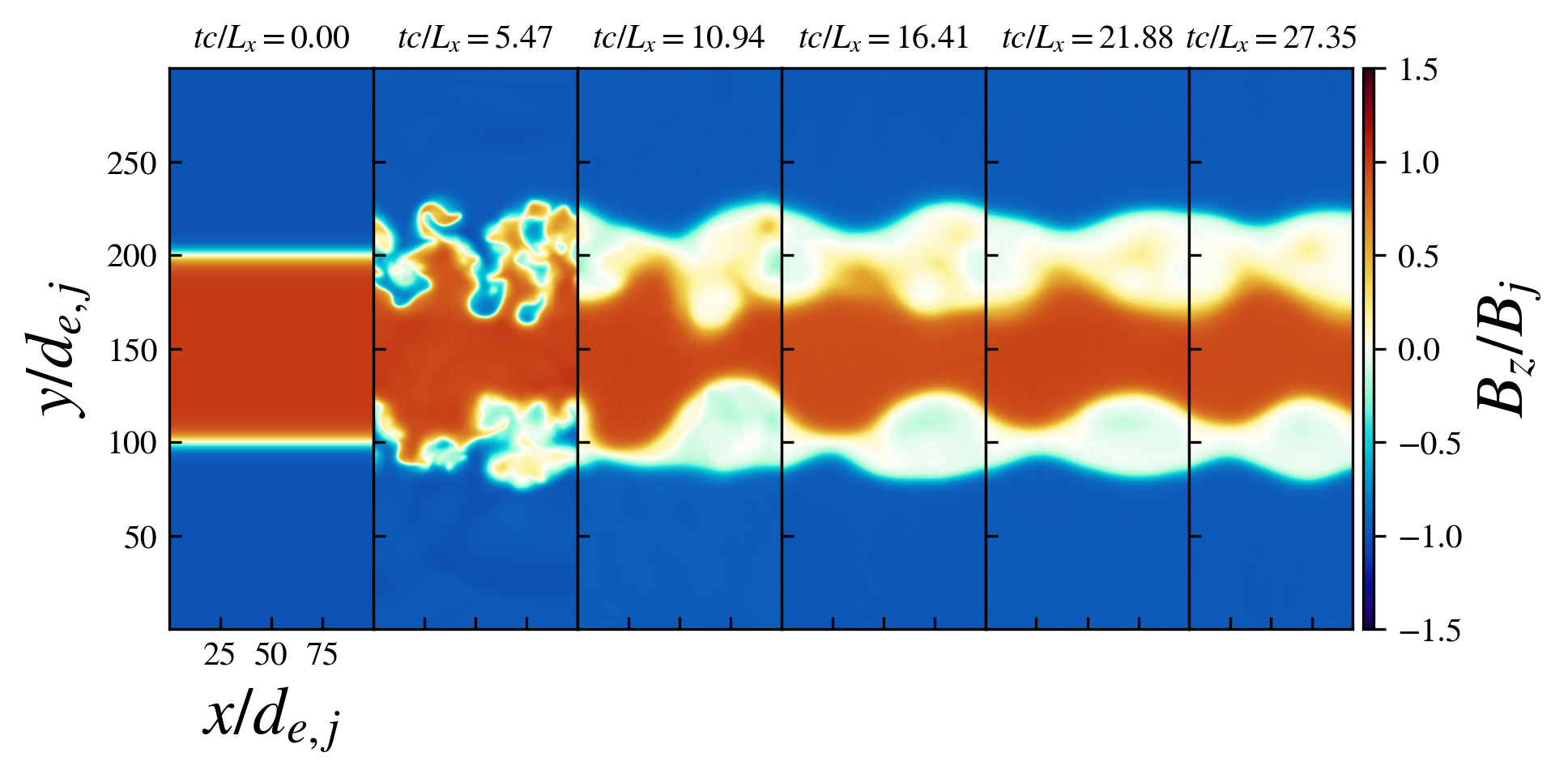} \\
    \includegraphics[width=0.7\textwidth]{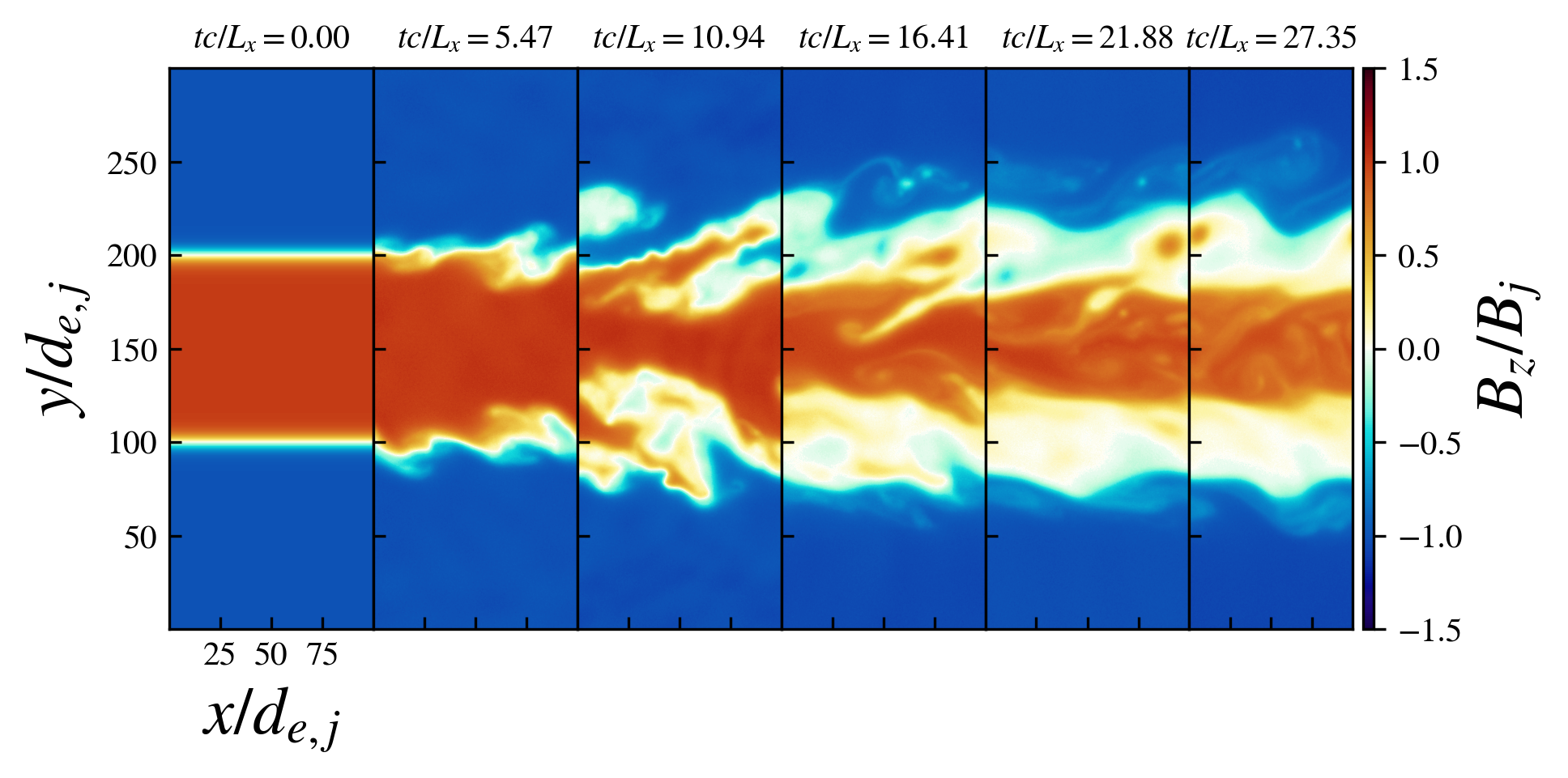} \\
    \includegraphics[width=0.5\textwidth]{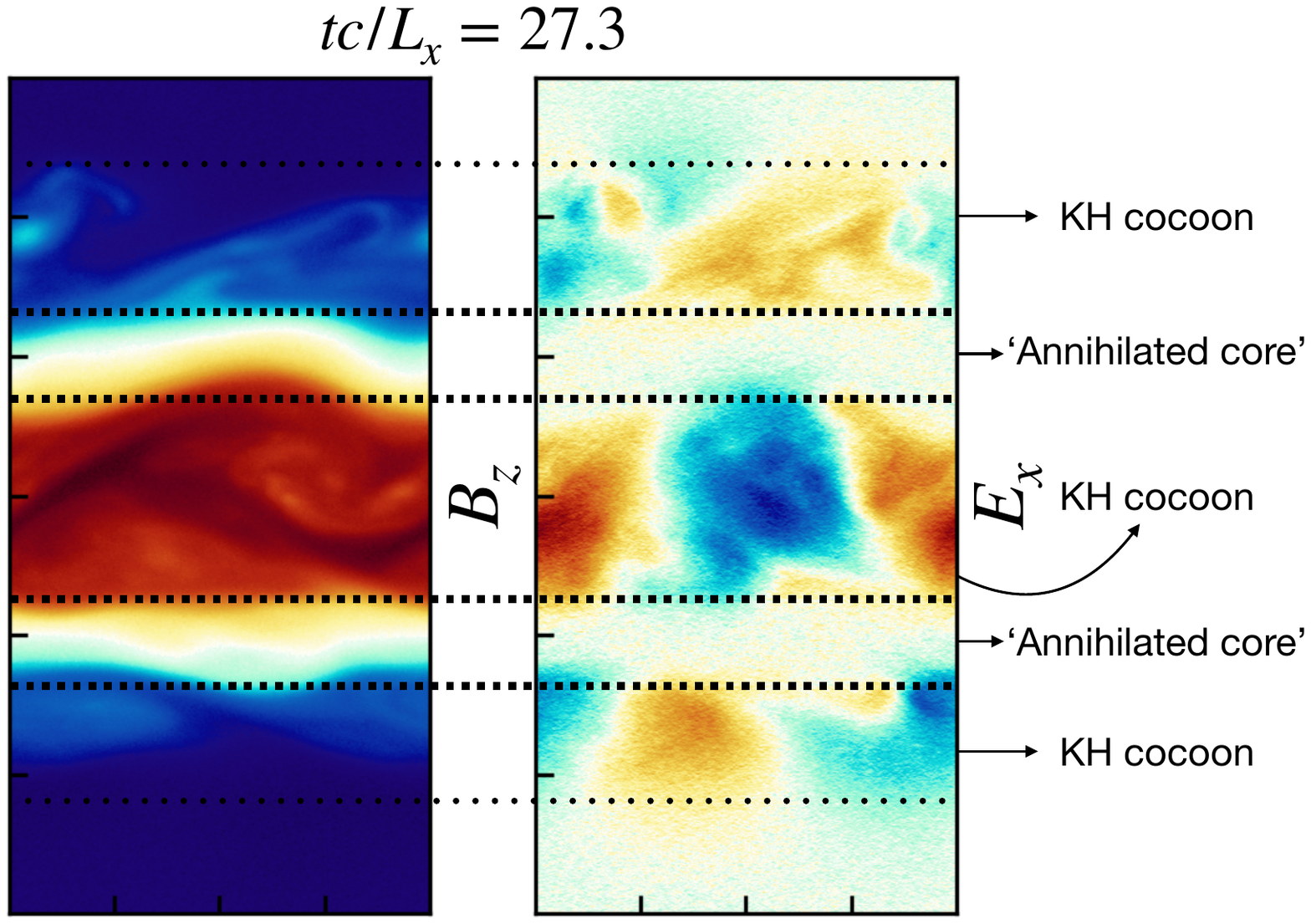}
    \caption{Snapshots of $B_z/B_j$ for simulations with velocity shear only (`VS', upper-top), magnetic shear only (`MS', upper-middle), and both shears ($u_j=0.5,B_w/B_j=-1$, upper-bottom). The lower panel shows snapshots of $B_z, E_x$ in the saturated stage ($tc/L_x=27.3$) for the $u_j=0.5,B_w/B_j=-1$ case, with the `annihilated core' and `KH cocoon' annotated.}
    \label{fig:snap_Bz}
\end{figure}

\section{Results} \label{sec:results}

\subsection{Morphology and dissipation} \label{subsec:morp_diss}

In this section, we first give a quick overview of the evolution of selected simulations, summarizing the stages they go through; then we discuss each stage in greater detail, focusing on the linear growth rate, nonlinear shear-layer width, and dissipation.

Fig.~\ref{fig:snap_Bz} shows the typical evolution of our simulations, which first follows a linear stage, then a nonlinear stage when the conversion of magnetic or bulk kinetic energy into internal energy takes place, and finally a saturated stage. The onset times of these stages are different for the~MS, mixed-shear, and VS cases. In the MS case, the linear stage occurs at $0<tc/L_x\lesssim 2.5$, the nonlinear stage at $2.5\lesssim tc/L_x\lesssim13$, and the saturated stage from $tc/L_x\gtrsim 13$. For the mixed-shear cases, the linear stage ends roughly at $tc/L_x\approx 2.5$, as in the MS case, but the nonlinear stage is somewhat prolonged, lasting until $tc/L_x\approx 25$ before transitioning to the saturated stage. For the VS case, the linear stage ends at $tc/L_x\approx 7$, the nonlinear stage lasts until $tc/L_x\approx 13$, and is then followed by the saturated stage. The upper two panels of Fig.~\ref{fig:snap_Bz} show the evolution of the two control cases: the velocity-shear-only (VS) case ($B_j/B_w = 1$), where only the KHI is excited, and the magnetic-shear-only (MS) case ($u_j=0$), where only the DKI is excited. The instabilities give rise to prominent features in the nonlinear stage: cat's-eye vortices can be clearly seen in the VS case (upper-top panel of Fig.~\ref{fig:snap_Bz}), while Rayleigh-Taylor-like plume features (upper-middle panel of Fig.~\ref{fig:snap_Bz}) arise in the MS case. The nonlinear phase begins when the $y$ displacement of the features becomes comparable to the modal wavelength in the $x$-direction. 
For the VS case this is characterized only by a slight smearing of the KH vortices, while the DK plumes mix together violently. This is the stage where most of the dissipation (if any) occurs. 
In the later saturated stage, the KH vortices persist to the end in the VS case, while in the MS case the mixing leads to a thickened shear layer where the electric and magnetic fields are drastically reduced. The DK plumes do not persist to the end.

When both types of shear are present, the KH and DK instabilities interact. The nonlinear stage ($tc/L_x\gtrsim 5-6$) is marked by the nonlinear interactions of the KH vortices and DK plumes, creating a turbulent shear layer (e.g. upper-bottom panel of Fig.~\ref{fig:snap_Bz}) at around $7-10 L_x/c$ (a small multiple of the DK growth timescale). As discussed below, the dissipation level is also modified significantly. In the saturated stage, the turbulence subsides, giving way to a shear layer consisting of a broad `annihilated core' where the $E$ and $B$ fields are suppressed, the flow is nearly stagnant, and the thermal pressure dominates. This relatively quiet layer is enshrouded by an active `KH cocoon' (lower panel of Fig.~\ref{fig:snap_Bz}).

\begin{figure}
    \centering
    \includegraphics[width=0.45\textwidth]{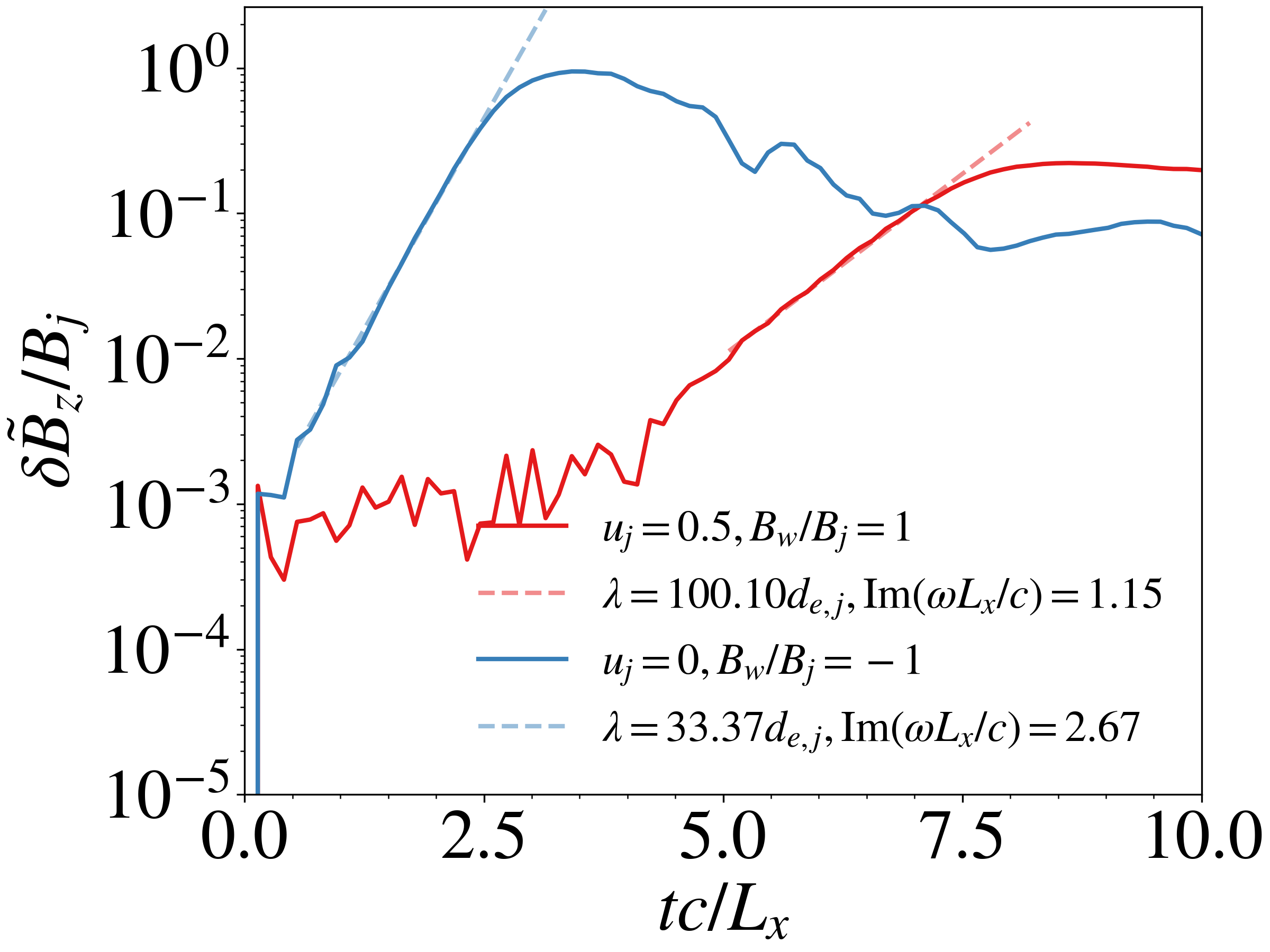}
    \includegraphics[width=0.45\textwidth]{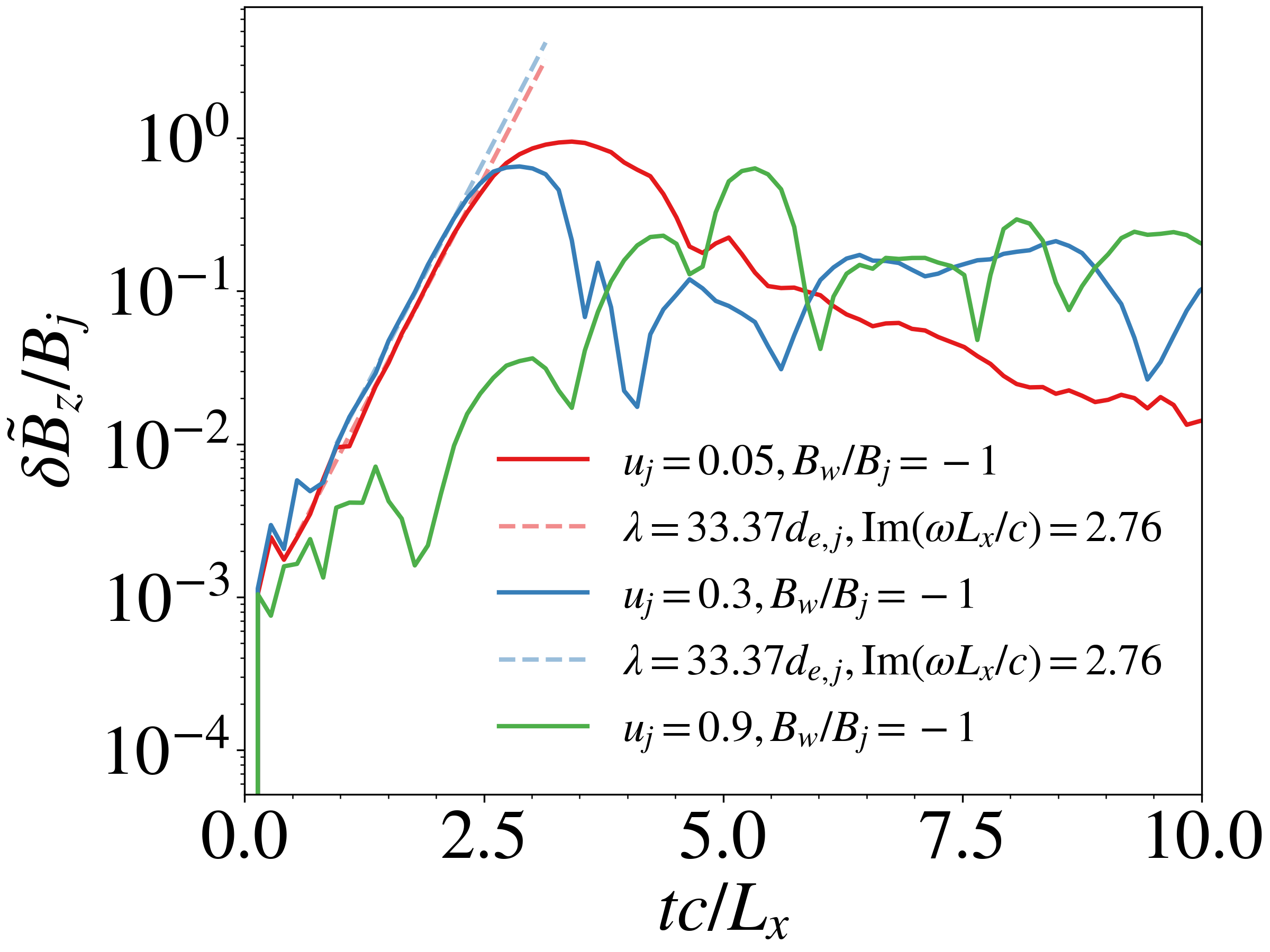} \\
    \includegraphics[width=0.45\textwidth]{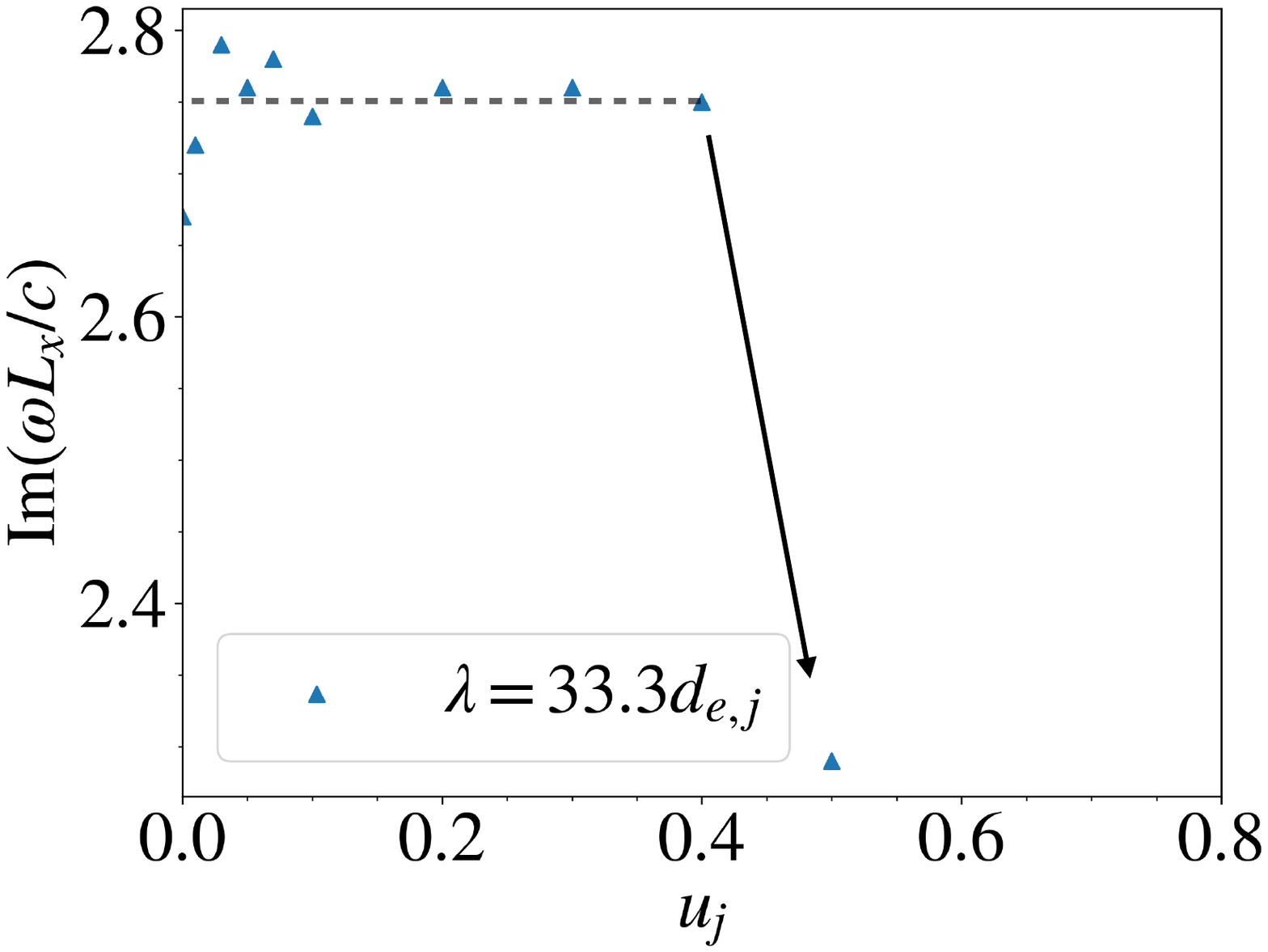}
    \caption{Top left: Growth curves of selected modes (indicated by the wavelength $\lambda$) for the control cases (red for VS, blue for MS). Note that the wavelengths of the displayed modes are different for the MS ($\lambda = 33 d_{e,j}$) and VS ($\lambda = 100 d_{e,j}$) case.  Top right: Growth curves for selected test cases (red for $u_j=0.05$, blue for $u_j=0.3$, green for $u_j=0.9$). In both panels, dashed lines show fitted linear growth rates as indicated in the legends. The growth curve of the $u_j=0.9$ case is not fitted as no exponential growth phase can be identified. Bottom panel: Fitted linear growth rates for the test cases, with the fastest growing mode displayed in the legend. No fitted growth rate is displayed for the $u_j>0.6$ cases as no exponential growth phase can be identified. Not that the growth rate of the VS case is not displayed in this panel. The black arrow highlights the drop in growth rate for the $u_j=0.5$ case.}
    \label{fig:linear_growth}
\end{figure}

To examine how velocity shear modifies the linear growth rate of the~DKI, we calculate the deviation of the $z$-component magnetic field from the initial value, i.e. $\delta B_z =B_z(x,y,t) - B_z(x,y,0)$, at $y/d_{e,j}=200$ (which is the initial location of the upper shear layer) and then Fourier-transform it in the $x$-direction, i.e. compute $\delta\tilde{B}_z(k_x, t)$. We then select for our analysis the Fourier mode whose amplitude first reaches $\delta\tilde{B}_z/B_j=0.1$ (which is typically $\lambda\approx 33d_{e,j}$ for the MS and mixed-shear case and $\lambda\approx 100d_{e,j}$ for the VS case, or equivalently $k_x\Delta\sim 1$ where $\Delta$ is the initial shear-layer width, as shown in Fig.~\ref{fig:linear_growth}). The variation of $\delta\tilde{B}_z/B_j$ of this mode with time (in units of $tc/L_x$) is displayed in Fig.~\ref{fig:linear_growth}. We observe, for most test cases (VS, MS, and the $u_j<0.6$ cases), a clear linear phase, exponential growth within the first~$2.5 L_x/c$. For the VS and MS cases (top left panel), two very different growths can be seen, corresponding to the KH-only (red line, VS) and DK-only (blue line, MS) instabilities. Using equation~25 in \citet{Zenitani_Hoshino-2007} to estimate the growth rate of the relativistic Drift-Kink instability in the MS case, we find an analytic expectation of $\mathrm{Im}(\omega L_x/c)\approx 3.05$, fairly consistent with our measured simulation result $\mathrm{Im}(\omega L_x/c)\approx 2.67$\footnote{The analytic growth rate of DKI used in this comparison assumes a two-fluid model, an adiabatic index of  $4/3$, a mode wavelength long compared to the initial shear-layer thickness, and negligible gyro-viscosity (see \citealt{Daughton-1999} for why it could be important). In fact, \citet{Zenitani_Hoshino-2007} found that the analytic formula overestimates the simulated growth rate as well, by a margin consistent with what we observe.}. Adopting the linear analysis described in \citet{Chow_etal-2023}, we calculate the analytic growth rate for the KH instability to be $\mathrm{Im}(\omega L_x/c)\approx 2.47$ for the VS case, whereas our simulation result gives $\mathrm{Im}(\omega L_x/c)=1.15$, which is a factor of 2 lower. Possible reasons for this discrepancy are noted in the footnote\footnote{The analytic growth rate of the KH instability used in this comparison assumes a razor thin transition layer, an adiabatic index of $4/3$, and the relativistic MHD equations \citep{Chow_etal-2023}. It is known that the KHI growth rate evaluated from a fluid model agrees well with the collisionless KHI growth rate as long as the gyroradious is much shorter than the wavelength. However, the growth rate evaluated from a razor thin shear-layer model typically overestimates the actual growth rate in simulations with finite shear layer thickness. Thus our rather low measured KHI growth rate is not unexpected.}. Even though the KH and DK modes separately grow on different time scales and length scales, as shown in the top left panel for the VS and MS cases, there is only one linear, exponential-growth phase in the mixed-shear cases ($u_j\neq 0,B_w/B_j=-1$), as shown in the top right panel (e.g. red and blue lines). From Fig.~\ref{fig:linear_growth}, the linear stage for this mode lasts for $2.5 L_x/c$ for the MS case and all the mixed-shear cases with $u_j<0.4$, whereas for the VS case it lasts until about $7 L_x/c$. When $u_j$ is increased above $0.5$, no exponential growth phase can be identified. 

We extract the growth rates by fitting the linear phase whenever it can be identified and plot the results as a function of $u_j$ in the bottom panel of Fig.~\ref{fig:linear_growth}. The wavelength of the displayed mode is $\lambda=33.3d_{e,j}$ (note that only the MS and mixed-shear cases are displayed in this panel).
As shown in this panel, the growth rates of all cases with an identifiable linear growth phase  (except the $u_j=0.5$ case) are roughly the same [$\mathrm{Im}(\omega L_x/c)\sim 2.75$, equivalently $\mathrm{Im}(\omega/\omega_{pe,j})\sim 0.0275$ and $\mathrm{Im}(\omega/\omega_{ce,j})\sim 0.0138$], equal to that in the MS case. This, together with the fact that the fastest growing modes for the mixed-shear cases are the same as in the MS case, implies that the DK modes still dominate the linear growth in the presence of moderately strong velocity shear ($u_j\lesssim 0.4$), and there is minimal interaction of the DK and KH modes in this linear stage. The effect of velocity shear in the linear phase becomes more apparent for $u_j\geq 0.5$ --- the growth rate drops to $\mathrm{Im}(\omega L_x/c)\sim 2.29$ for $u_j=0.5$ and no exponential growth phase can be identified for stronger shears.

\begin{figure}
    \centering
    \includegraphics[width=0.45\textwidth]{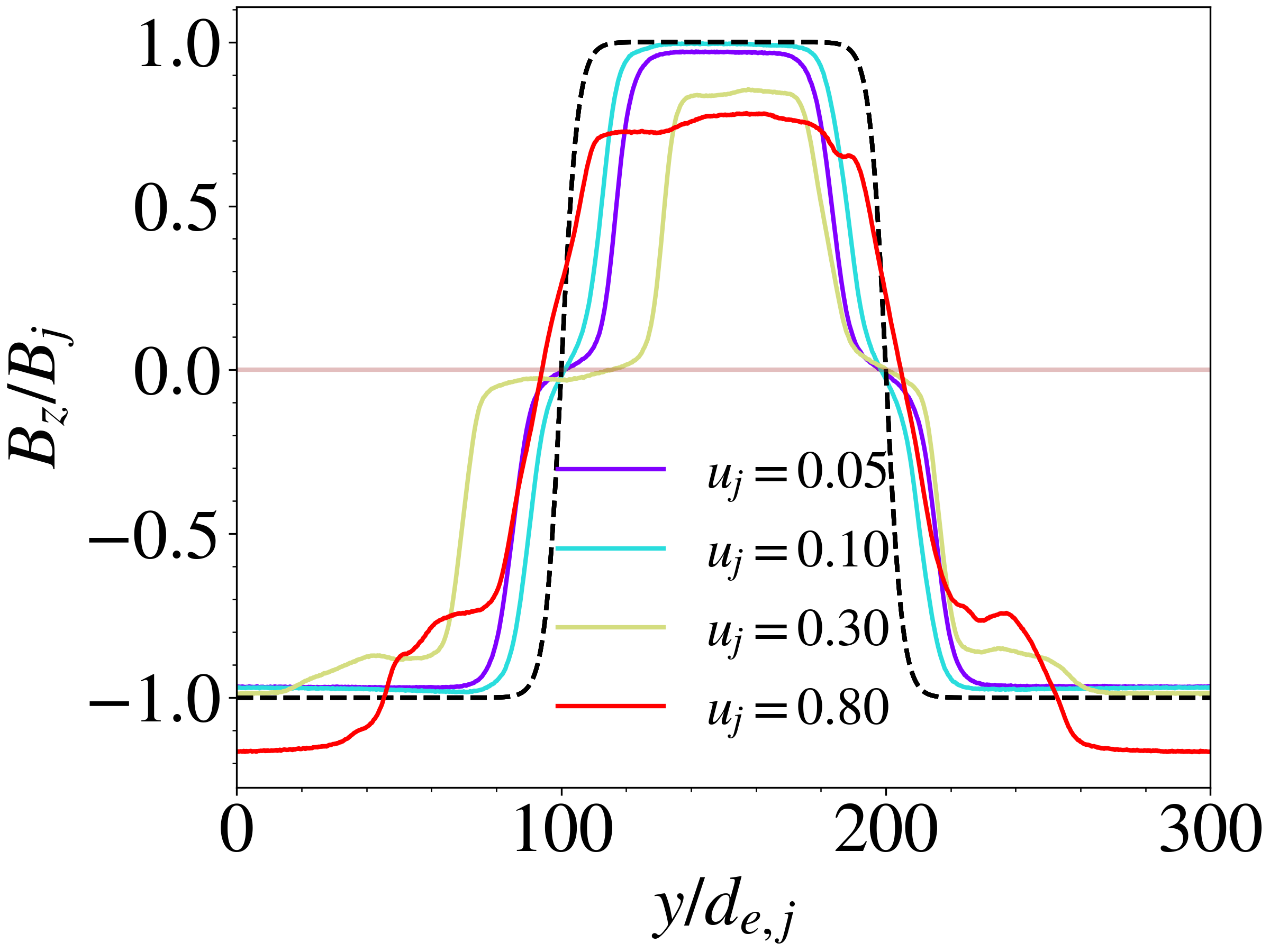} 
    \includegraphics[width=0.45\textwidth]{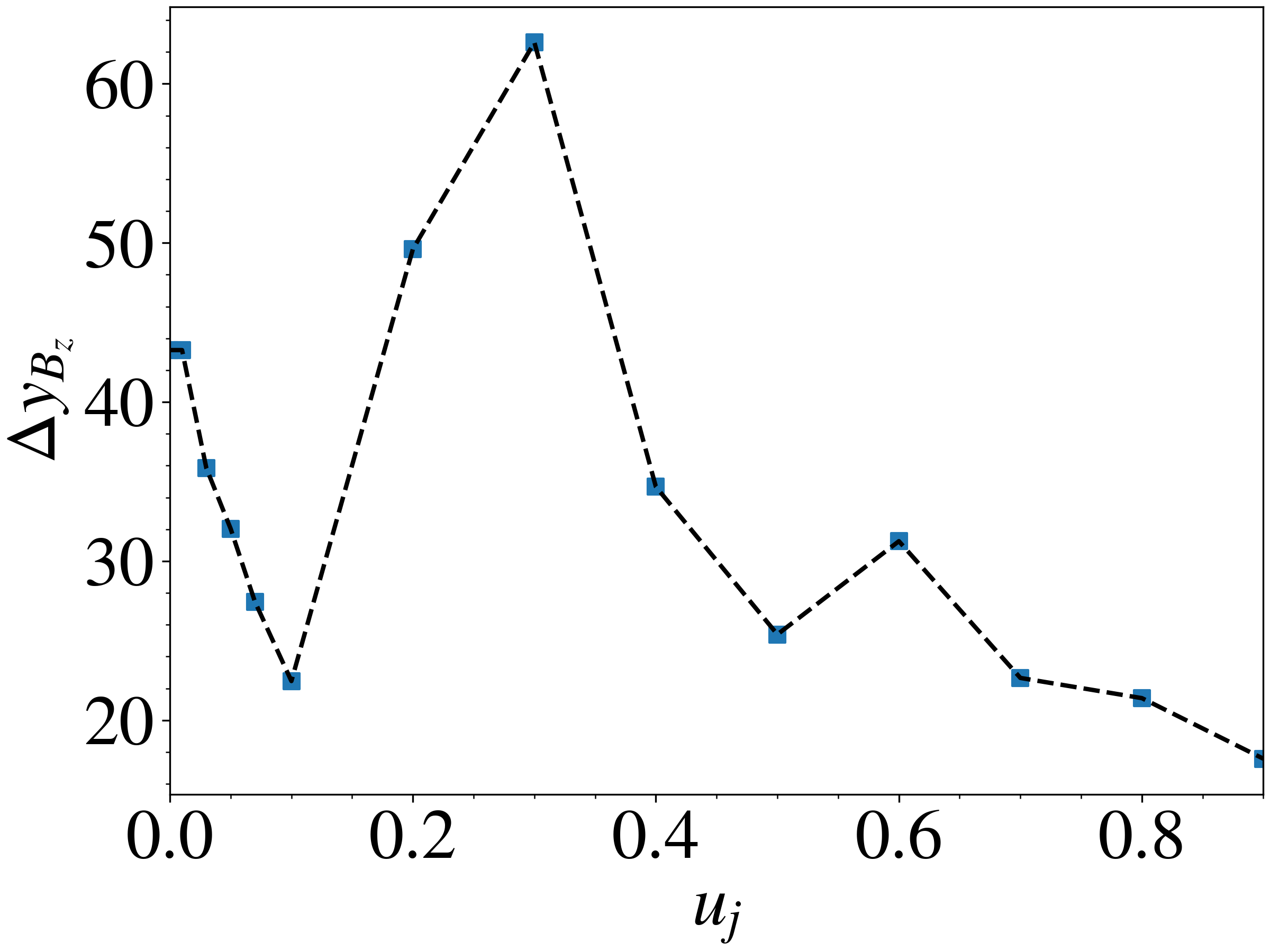}
    \caption{
    Left: $x$-averaged plots of $B_z/B_j$ for $u_j = 0.05,0.1,0.3,0.8$, $B_w/B_j = -1$, showing how the instabilities thicken the magnetic shear layer.
    The black dashed line is the $x$-averaged plot of $B_z/B_j$ at $t=0$. The horizontal translucent brown line indicates $B_z=0$. Right: Width of the thickened shear layer as a function of velocity shear, at $tc/L_x=55$.}
    \label{fig:growth}
\end{figure}

Despite the relatively unchanged linear growth rates observed at early times for $u_j\lesssim 0.4$ cases, the morphological changes at late times are apparent. In the following, we characterize the structure of the test cases by the width of the thickened shear layer. On the left panel of Fig.~\ref{fig:growth} we plot the $x$-averaged profiles of $B_z/B_j$ at the final time $tc/L_x = 55$ for several values of~$u_j$. We observe that turbulence driven by nonlinear interactions of the instabilities has thickened the shear layer, within which $B_z$ is reduced. We measure the width of this layer $\Delta y_{B_z}$ at $tc/L_x=55$ by the $y$-separation between isosufaces of $B_z/B_j = -0.5$ and $0.5$ and display it as a function of~$u_j$ in the right panel of Fig.~\ref{fig:growth}. A nontrivial picture emerges. The width is substantial in the pure-DKI MS case ($u_j=0$), but decreases with increased shear for relatively weak $u_j\lesssim 0.1$. It then bounces back up for intermediate values of $u_j$ ($0.1<u_j<0.3$), but drops again for larger values of~$u_j$. This is quite different from the $u_j$-dependence of the linear growth rate, which shows no such variations for $u_j\lesssim 0.4$. We therefore attribute the changes in the shear-layer width to nonlinear interaction of the instabilities. Namely, when the velocity shear is moderately weak ($u_j<0.1$), DK plumes protruding across the plasma's bulk flow are quickly sheared away, reducing the layer's width. As $u_j$ increases to around~0.2, KHI kicks in and the cat's-eye vortices wrap up the layer, causing it to thicken again. When the velocity difference becomes super-magnetosonic, however, KHI is suppressed \citep{Chow_etal-2023}, leading to a thinner saturated layer. We observe that the onset of KH vortices, if it happens, occurs at a slightly later time than the linear, DK-dominated growth phase. For example, in the $u_j=0.3$ case, the KH vortices set in at around $tc/L_x\sim 5$, whereas the linear phase occurs between $0< tc/L_x< 2.5$ (observe in the top right panel of Fig.~\ref{fig:linear_growth} that there is a small increase in $\delta\tilde{B}_z/B_j$ at $tc/L_x\sim 5$ for the blue line, indicative of KH vortex wrap-ups). As shown below, the width of the thickened shear layer correlates strongly with magnetic energy dissipation.

\begin{figure}
    \centering
    \includegraphics[width=0.45\textwidth]{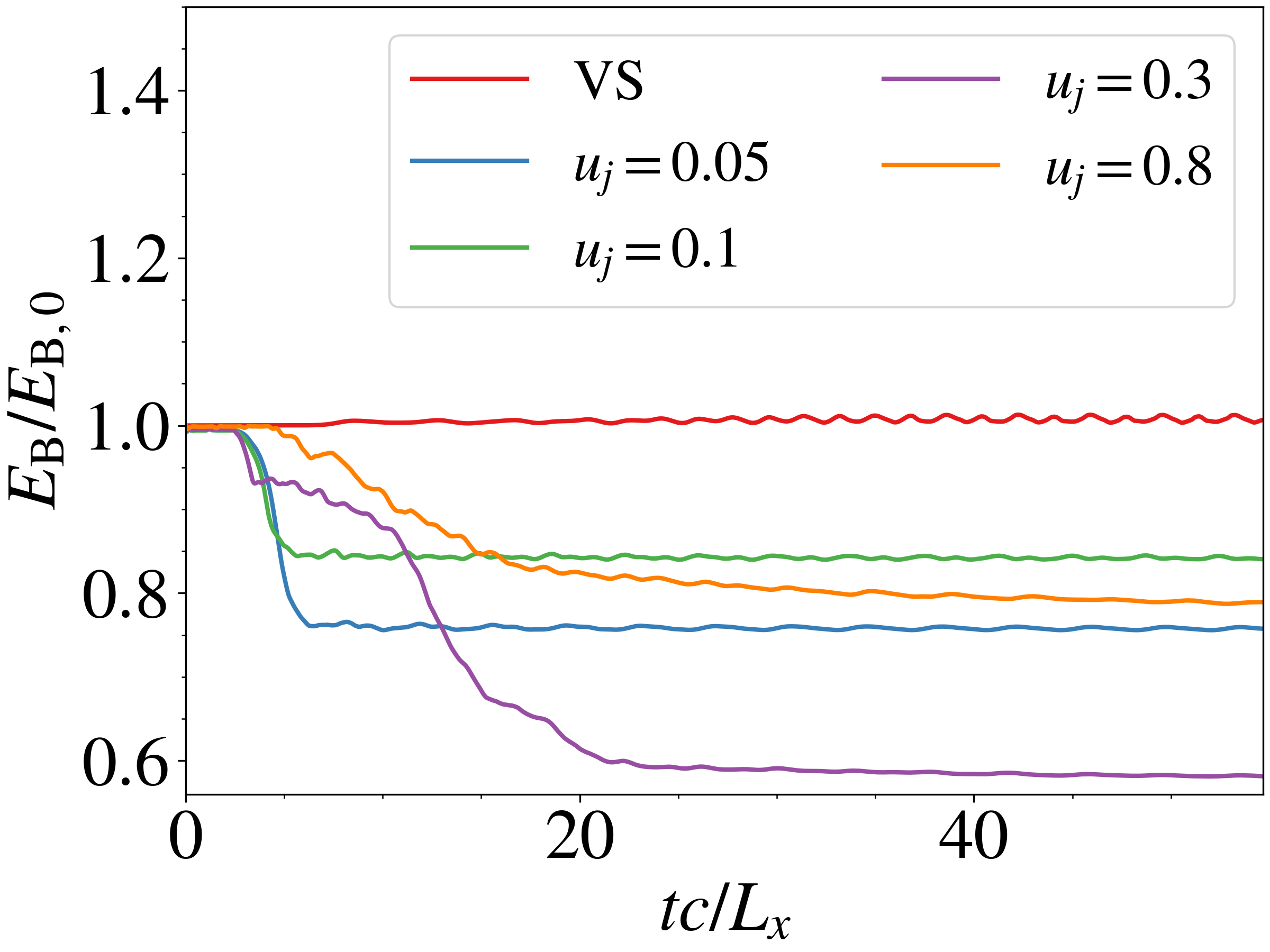}
    \includegraphics[width=0.45\textwidth]{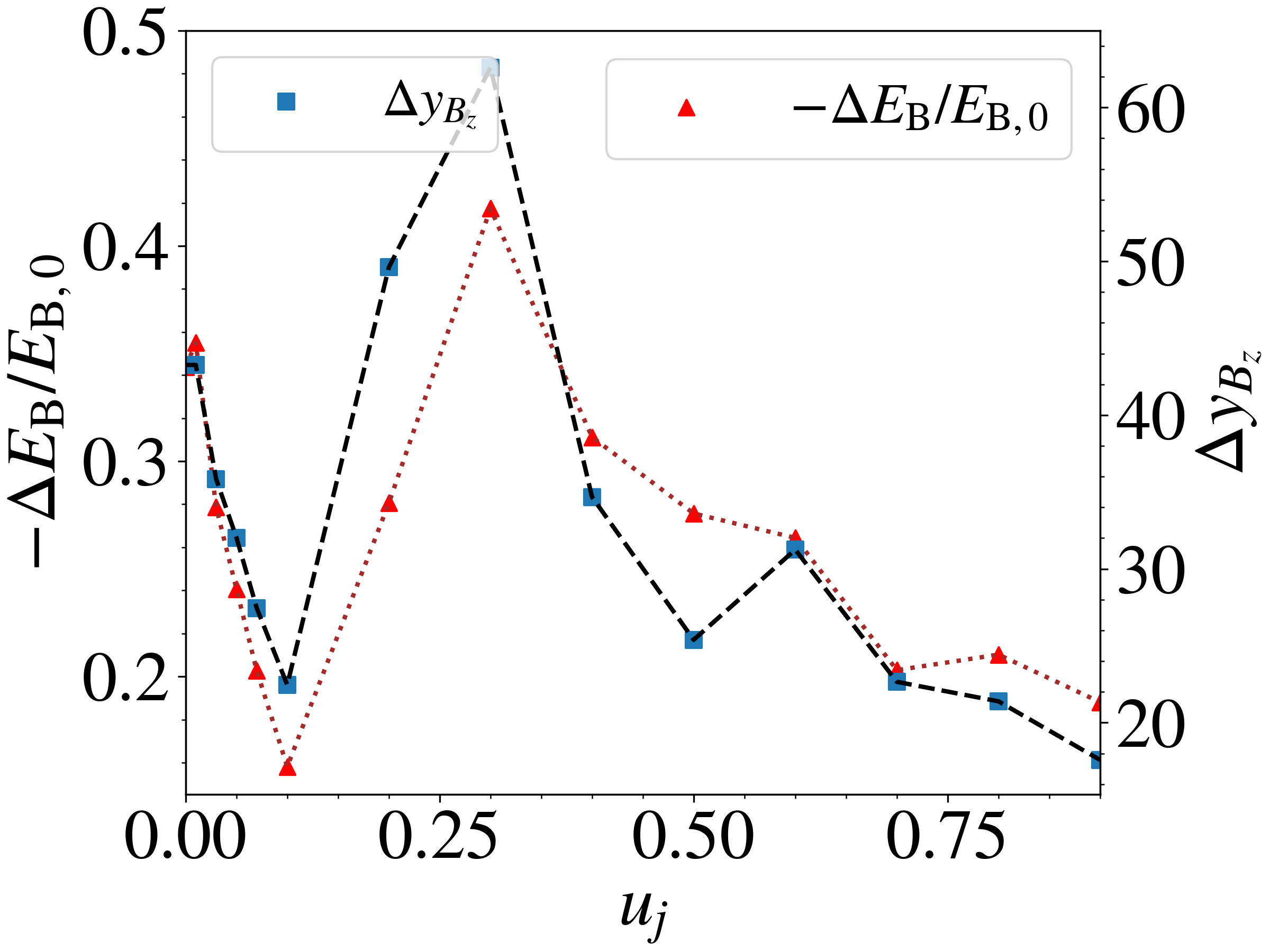}\\
    \includegraphics[width=0.45\textwidth]{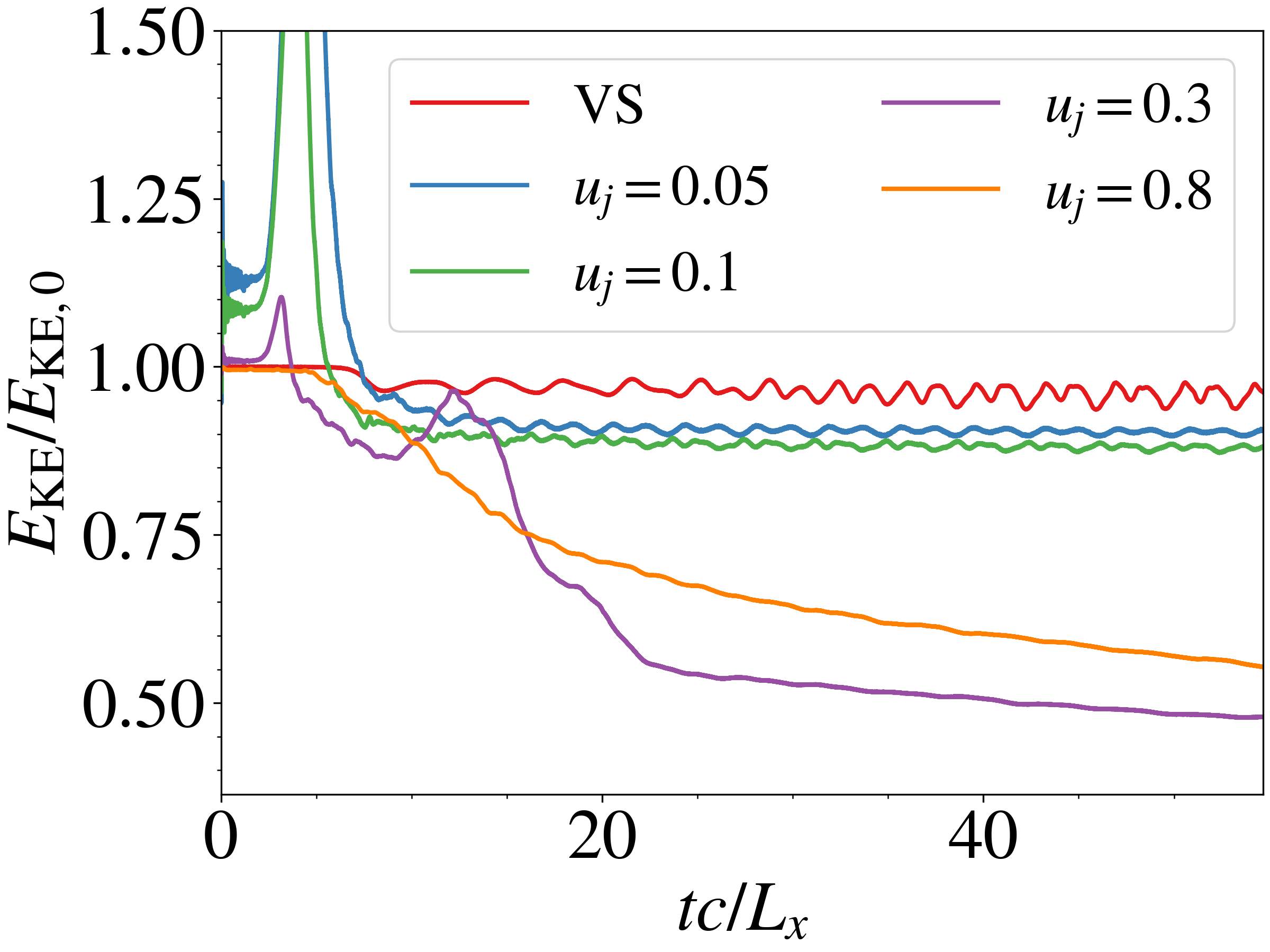}
    \includegraphics[width=0.45\textwidth]{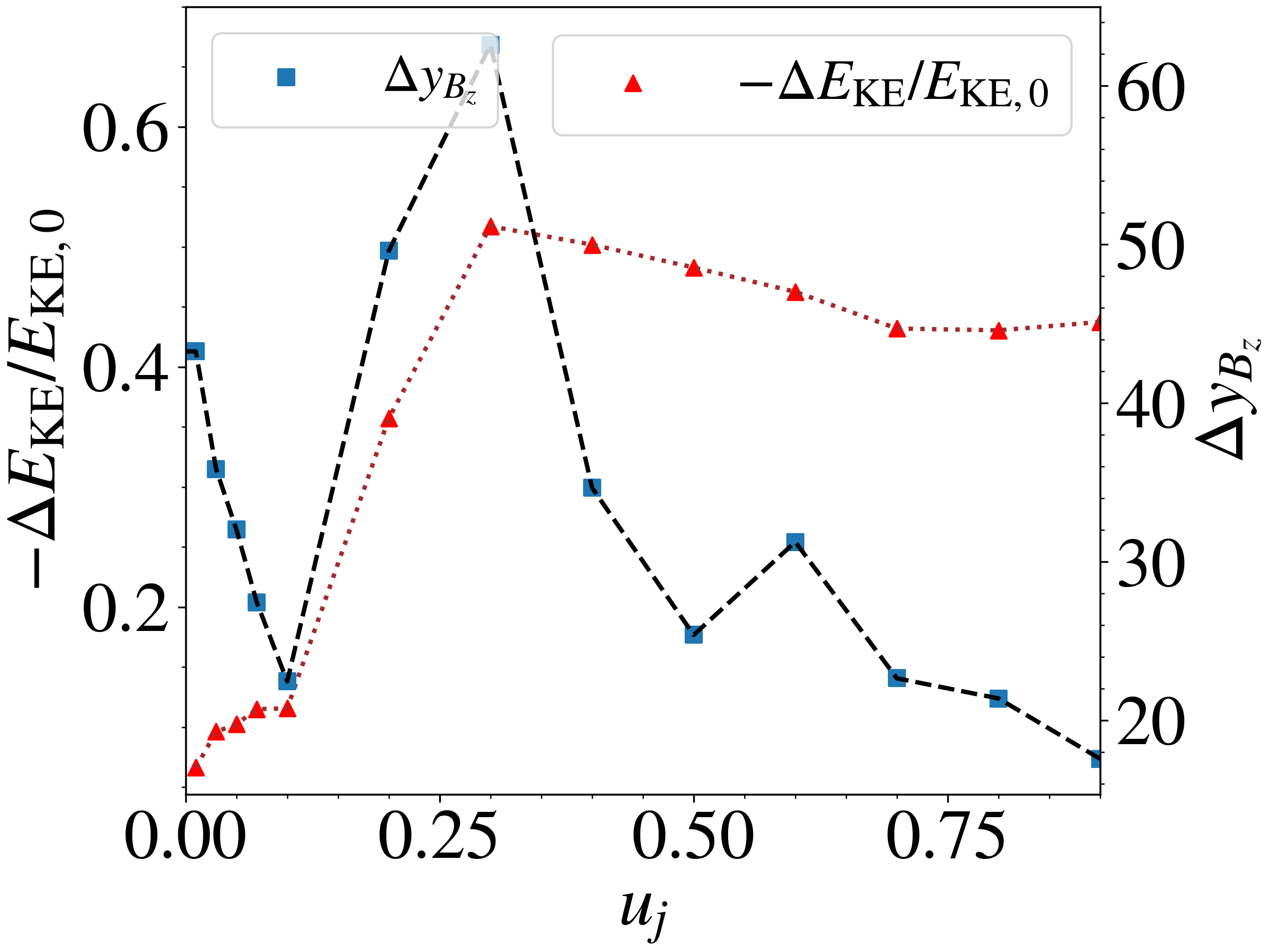}
    \caption{Top and bottom left: The total magnetic and bulk kinetic energies within the simulation box $E_B, E_\mathrm{KE}$, normalized by their initial values $E_{B,0}, E_\mathrm{KE,0}$, against time for selected cases ($u_j = 0.05,0.1,0.3,0.8$, $B_w/B_j = -1$ and the VS case). Top and bottom right: brown dotted lines with red triangles show the magnetic and bulk kinetic energy dissipated, measured by $-\Delta E_B/E_{B,0},-\Delta E_\mathrm{KE}/E_\mathrm{KE,0}$, as a function of velocity shear~$u_j$, at $tc/L_x=50$. Black dashed lines with blue squares show the width of the thickened shear layer for different~$u_j$, same as the bottom right panel of Fig.~\ref{fig:growth}, superimposed for comparison.}
    \label{fig:dissipate}
\end{figure}

The dissipation of magnetic and bulk kinetic energy is closely tied to the instabilities. In the top and bottom left panels of Fig.~\ref{fig:dissipate} we show the evolution of the total magnetic and bulk kinetic energies, $E_B$ and $E_\mathrm{KE}$, normalized by their initial values, for selected cases. Most of the dissipation occurs at around $tc/L_x\approx 10$, when the perturbations have fully grown, with $E_B$ and $E_\mathrm{KE}$ steadying out afterwards. Up to half of the initial $E_B$ and $E_\mathrm{KE}$ can be dissipated as a result of the instabilities, but as shown in the top and bottom right panels of Fig.~\ref{fig:dissipate}, this is very sensitive to the velocity shear~$u_j$. In fact, the dissipation of magnetic energy for different $u_j$ closely mimics the trend for the width of the thickened shear layer. The reason for this is simple: the shear-layer is devoid of magnetic flux density, whereas outside the layer it is roughly the same as the initial value, so the total decrease of magnetic energy is just due to dissipation of the field inside this layer. This is why the decrease in $E_B$ is directly proportional to the unmagnetized layer's thickness~$\Delta y_{B_z}$. The dissipation of bulk kinetic energy also shows a prominent peak at moderate velocity shears $0.2<u_j<0.4$, but unlike $E_B$, the dissipation of $E_\mathrm{KE}$ is less effective for weak velocity shears $u_j<0.1$. We note also that the case with only velocity shear (VS) exhibits zero magnetic dissipation and close to zero (less than $5\%$) bulk-kinetic energy dissipation. While KH vortices alone do not appear to be able to dissipate magnetic energy or bulk kinetic energy effectively in this 2D configuration (at least within the timescales we explored), and even a weak velocity shear is enough to reduce the dissipation of magnetic energy with DK plumes alone, their combined effects are strong. There are two noteworthy points here: firstly, that the dissipation correlates with the thickening of the shear layer, and secondly, that the instabilities can synergistically increase the dissipation. With two shears, which could source two different instabilities, it is possible for one type of shear to limit how big the perturbations arising from the other type can grow, thus reducing dissipation; e.g. velocity shear disrupts DK plumes. At the same time, it is also possible for one type of shear to enhance the dissipative effect of perturbations arising from the other type of shear, as in the case of magnetic shear influencing KH vortices. 

\begin{figure}
    \centering
    \includegraphics[width=0.8\textwidth]{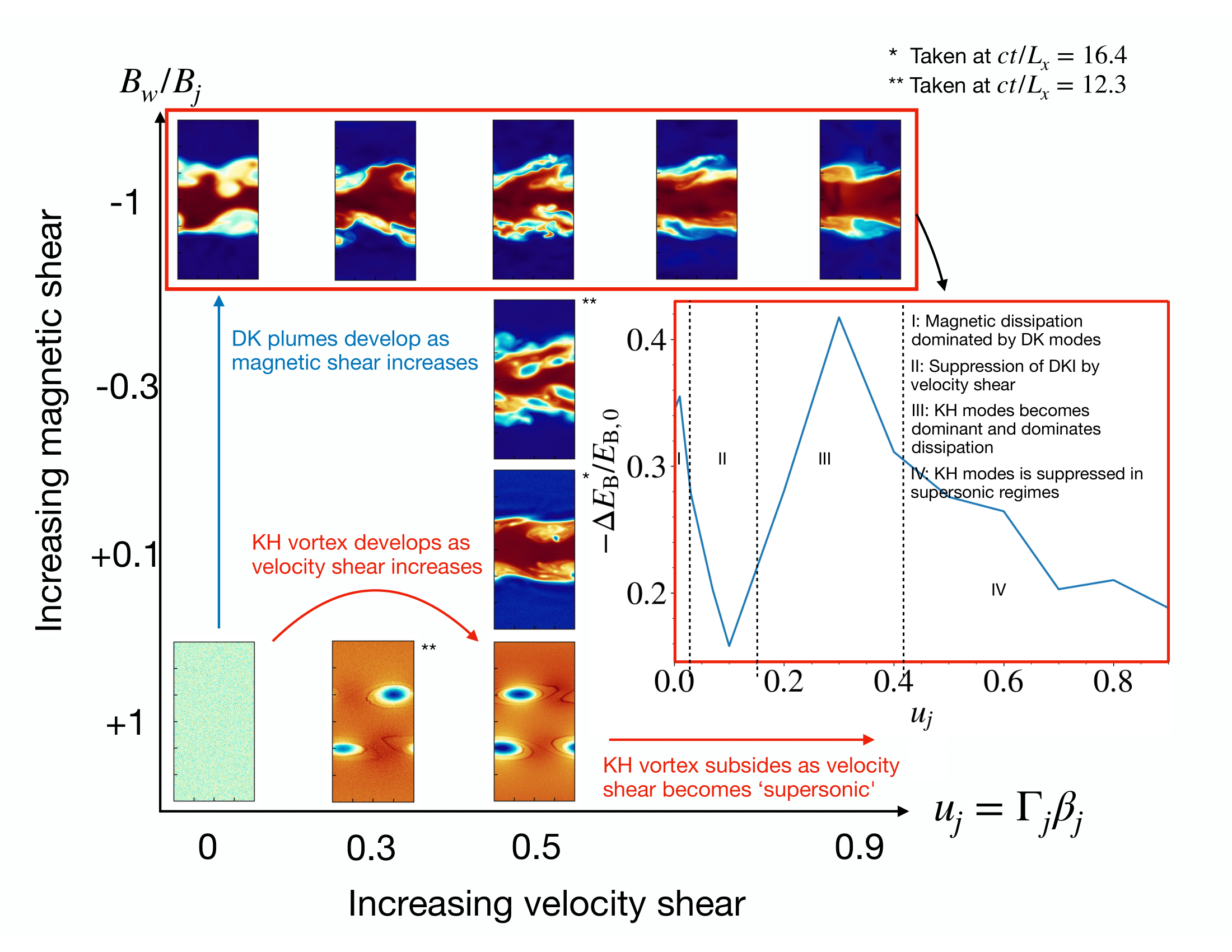}
    \caption{Schematic diagram showing how velocity and magnetic shear affect the plasma flow through the instabilities they excite, with the magnitude of the velocity shear (measured by~$u_j$, the `jet' 4-velocity) on one axis and the magnetic shear (measured by $B_w/B_j$) on the other. The snapshots are taken mostly at $tc/L_x=8.2$, with some taken at other times. The inset shows the fraction of magnetic energy dissipated as a function of $u_j$ (measured by $-\Delta E_B/E_{B,0}$, same as Fig.~\ref{fig:dissipate}) for the cases framed in red. The plot is divided into 4 regions, representing different regimes where a certain instability is active/suppressed.}
    \label{fig:schematic}
\end{figure}

We summarize our discussion up to this point with a schematic diagram (Fig.~\ref{fig:schematic}), showing how the flow structure and dissipation change with respect to the magnitude of the two shears we investigated, within the regimes we explored. As shown in the diagram, there are two free parameters, each controlling a type of shear, represented by the two axes. We show simulation snapshots for selected cases, each characterized by a combination $(u_j, B_w/B_j)$. In the absence of any shear $(u_j=0, B_w/B_j=+1)$, the result is just particle white noise. In the absence of magnetic shear (VS case; $B_w/B_j=+1$), KH vortices develop when the velocity shear increases, subsiding when $u_j$ becomes supersonic \citep{Chow_etal-2023}. In the absence of velocity shear (MS case; $u_j=0,B_w/B_j=-1$), Rayleigh-Taylor like plumes develop from DKI, developing into a thickened shear-layer with subdued electromagnetic field. Holding the velocity shear constant at $u_j=0.5$, we observe an increase in turbulent activity as the magnetic shear $B_w/B_j$ increases, indicating enhanced nonlinear interplay between the KH and DK instabilities. With the magnetic shear fixed at $B_w/B_j=-1$ while varying $u_j$, which is the subject of inquiry in this study, we observe the aforementioned change in morphology and dissipation. These changes can be divided into four regimes: I) without velocity shear, magnetic dissipation is dominated by the DK modes. II) In the presence of a weak velocity shear, the DK plumes are suppressed nonlinearly by the velocity shear, resulting in a drop in dissipation. III) With a moderately strong velocity shear, KH vortices are excited, resulting in nonlinear interplay with the DK modes, leading to enhanced dissipation. IV) With supersonic shear, KH subsides and dissipation reduces.

\subsection{Particle acceleration} \label{subsec:ptcl_acc}

\begin{figure}
    \centering
    \includegraphics[width=0.45\textwidth]{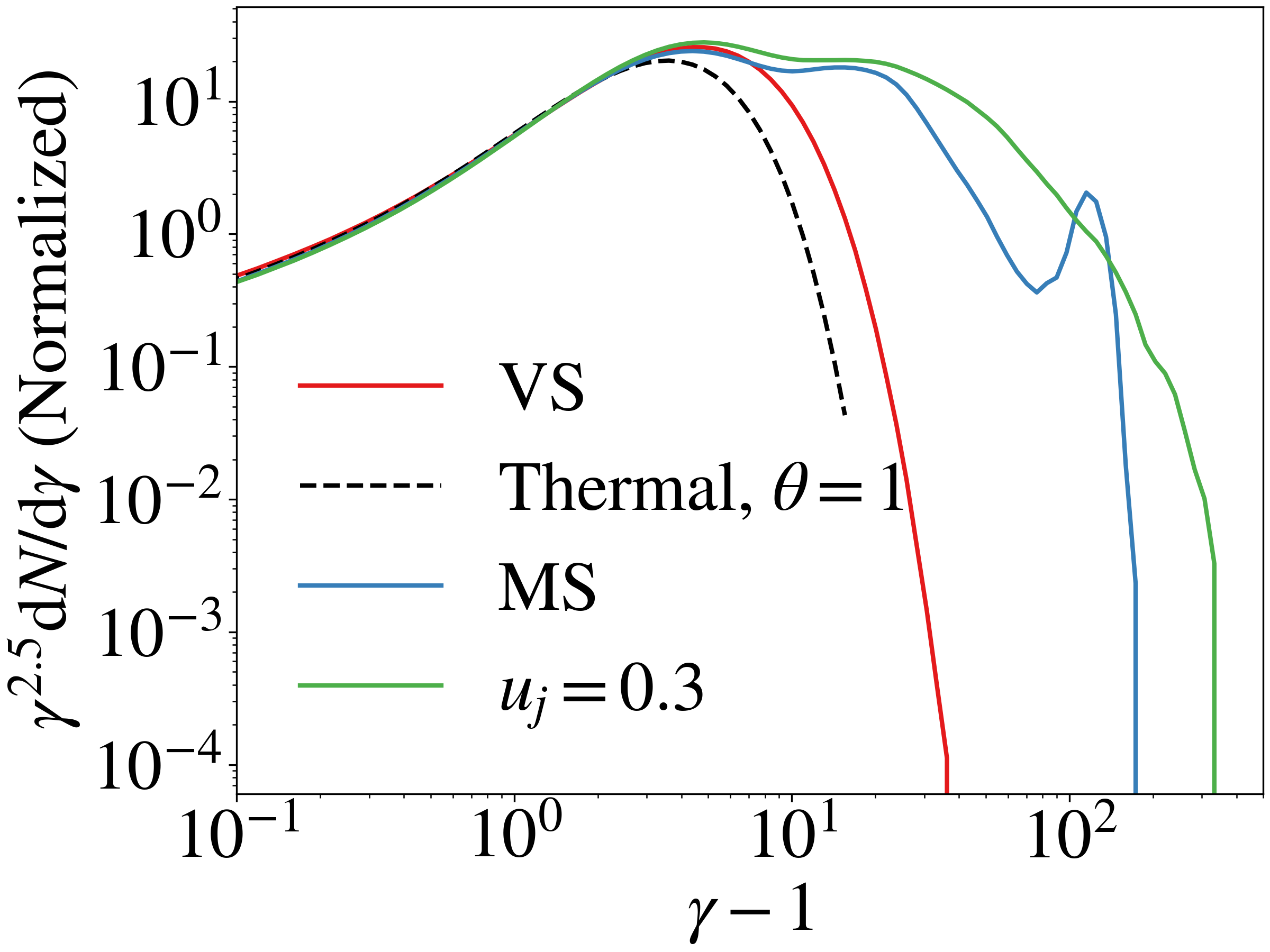}
    \caption{Particle energy distributions for the VS, MS and $u_j=0.3, B_w/B_j=-1$ cases at $tc/L_x=27.3$, showing the generation of a nonthermal power-law like tail. For comparison, a thermal (Maxwell-J\"uttner) distribution with temperature $\theta=1$ is shown with a black dashed line.}
    \label{fig:power_law}
\end{figure}

\begin{figure}
    \centering
    \includegraphics[width=\textwidth]{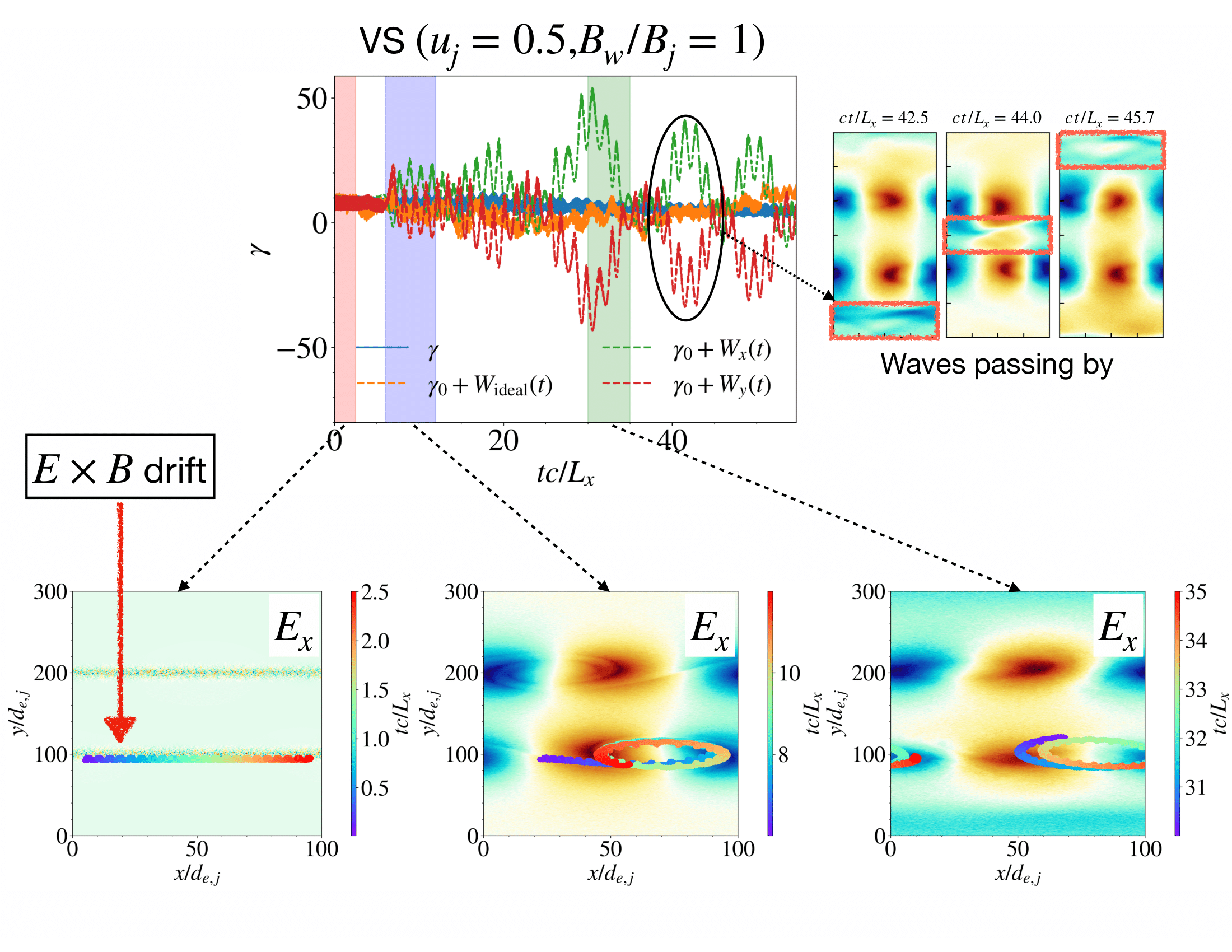} 
    \caption{Energetics and trajectory of a particle in the VS case. The selected particle has a Lorentz factor of $10$ at $tc/L_c=27.3$. Top panel: Lorentz factor $\gamma(t)$ as a function of time (blue), in comparison to the work done by various components of the $E$-field ($W_\mathrm{ideal}, W_x, W_y$, defined in the main text, representing the work done by the motional $E_x$ and $E_y$ fields). Bottom panels: Trajectories of the particle at three time windows (marked by red, blue and green in the top panel), overlaid on three $E_x$ snapshots taken at $tc/L_x=0, 8.75, 31.45$. The panel on the top right, showing three $E_x$ snapshots, displays the passage of waves (the passage of the wavecrest in the $y$-direction is highlighted by the red rectangular boxes), which generates the large scale bumps in the $E_x, E_y$ work done.}
    \label{fig:VS_track}
\end{figure}

\begin{figure}
    \centering
    \includegraphics[width=\textwidth]{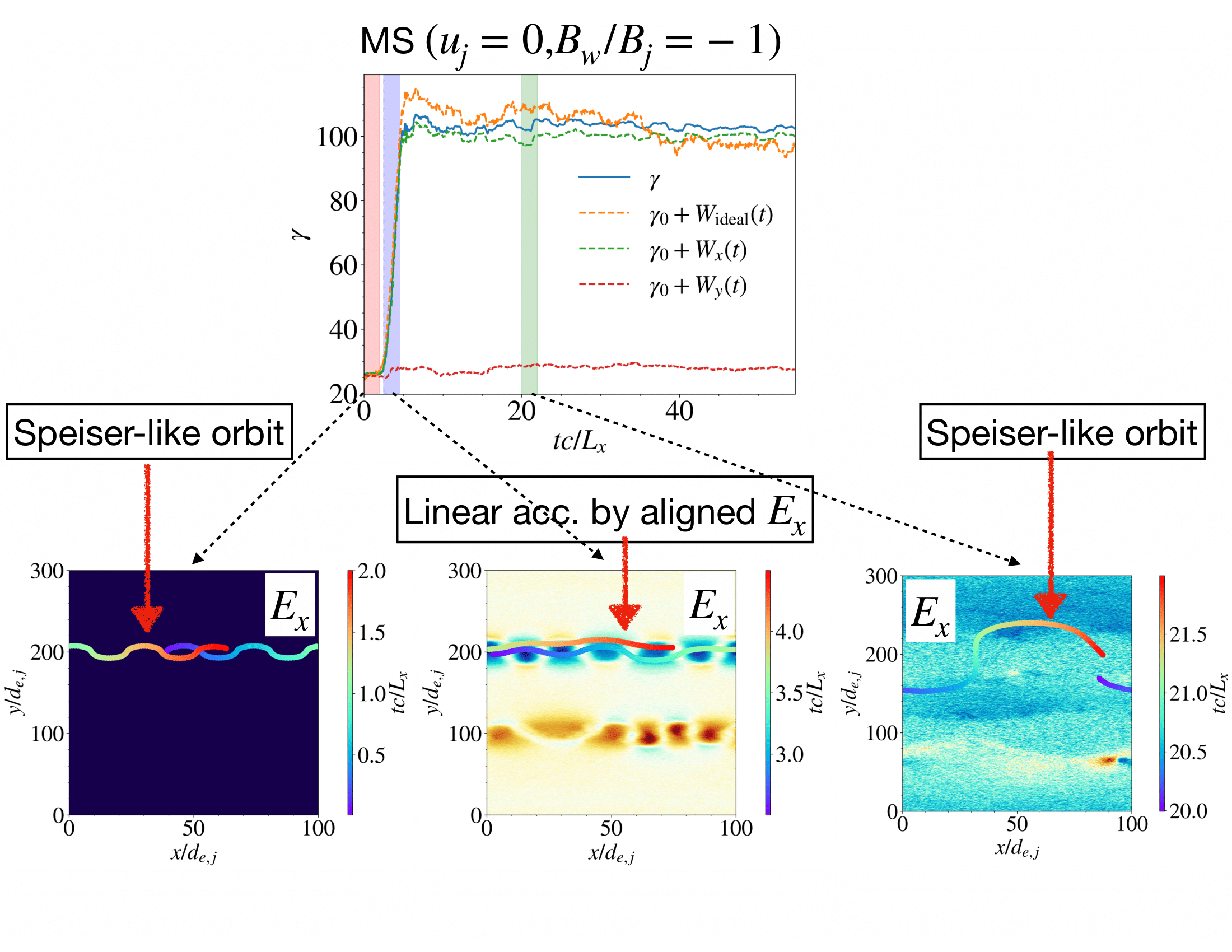} 
    \caption{Energetics and trajectory of a particle in the MS case. The selected particle has a Lorentz factor of $100$ at $tc/L_c=27.3$. The meaning of the plots and legends is the same as in Fig.~\ref{fig:VS_track}, with the difference that the trajectories in the bottom panels are taken at different times, and the $E_x$ snapshots in the background are taken at $tc/L_x=0, 3.85, 21.1$.}
    \label{fig:MS_track}
\end{figure}

\begin{figure}
    \centering
    \includegraphics[width=\textwidth]{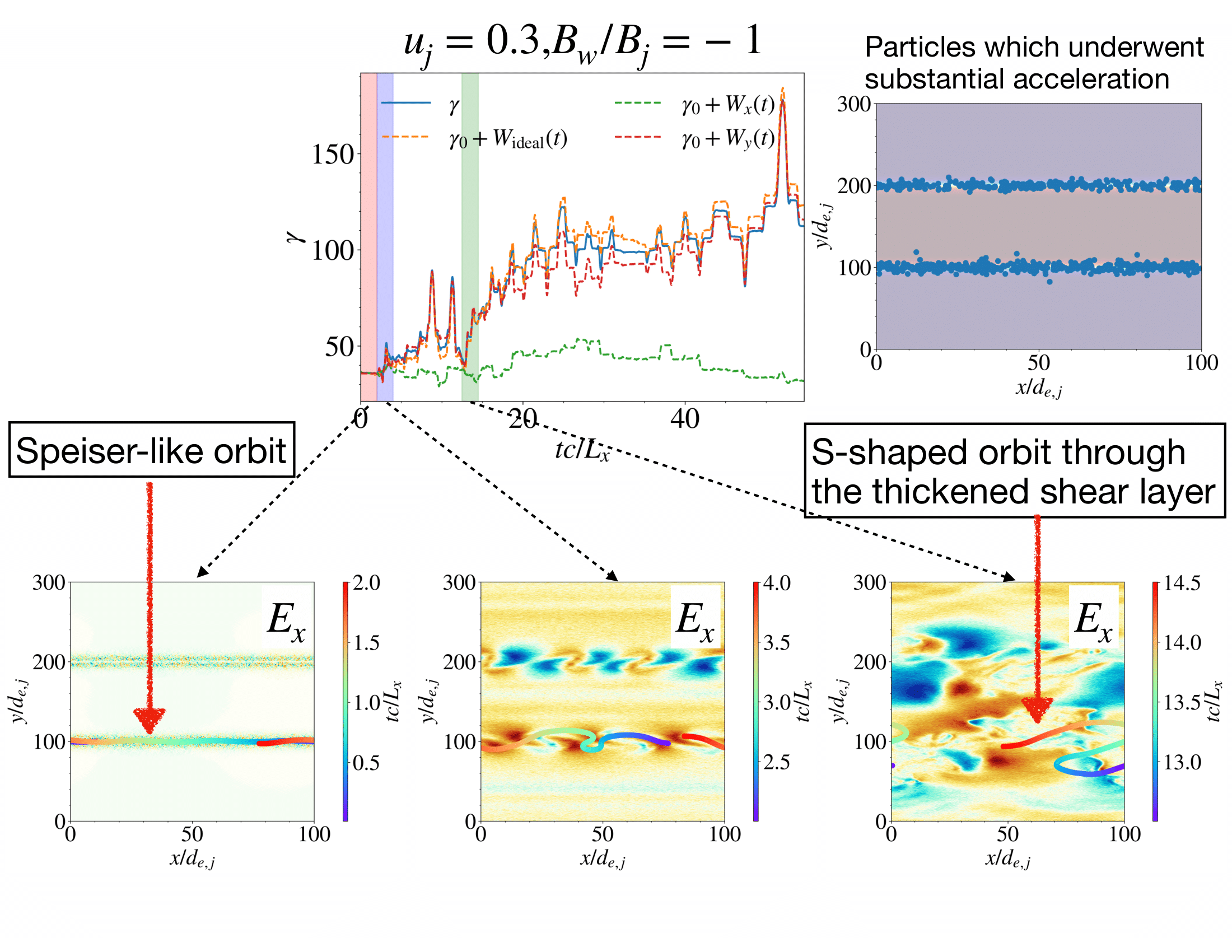}
    \caption{Energetics and trajectory of a particle in the mixed-shear ($u_j=0.3, B_w/B_j=-1$) case. The selected particle has a Lorentz factor of $100$ at $tc/L_c=27.3$. The meaning of the plots and legends are the same as in Fig.~\ref{fig:VS_track}, with the difference that the trajectories of the bottom panels are taken at different times, and the $E_x$ snapshots in the background are taken at $tc/L_x=0, 3, 15.5$. The top right panel shows the initial positions of particles (blue dots) that were accelerated to $\gamma>30$, superimposed on the initial $B_z$ background.}
    \label{fig:uj0.3_track}
\end{figure}

We now delve deeper into the kinetic aspects of the instabilities and dissipation by tracking particles that have been accelerated. We first point out the generation of a nonthermal tails in the mixed-shear $(u_j=0.3, B_w/B_j=-1)$ case in Fig.~\ref{fig:power_law}, showing a small power-law section from $3\leq\gamma\leq30$ with spectral index $\sim 2.5$ and a steeper spectrum from $30\leq\gamma\leq200$, suggestive of a scale-invariant dissipative process. We also observe a modest power-law tail at $3\leq\gamma\leq30$ with spectral index $\sim 2.5$ in the MS case, followed by a much steeper power law and then a pile-up of particles at $\gamma\sim 10^2$. The energy distribution of the VS case remains thermal. In in the top panels of  Figs.~\ref{fig:VS_track}-\ref{fig:uj0.3_track} we show, respectively for the VS, MS, and mixed $(u_j=0.3, B_w/B_j=-1)$ cases, the Lorentz factors of selected electrons as a function of time in comparison with the work done on them by various components of the electric field; three bottom panels in each case display the trajectories of the particles during three time windows, taken in the linear, nonlinear and saturated stages. Additional explanatory panels are also inserted. In the top panel of each figure, the blue solid line denotes the Lorentz factor $\gamma(t)$ of the particle, while the orange dashed line denotes the work done by the ideal-MHD, motional electric field, $W_\mathrm{ideal}(t) = -e\int_0^{t}\vb{E}_\mathrm{ideal}\cdot\vb{v}\dd{t'}/m_e c^2$; the green dashed line denotes work done by the $E_x$-component, $W_x(t) = -e\int_0^{t}E_x v_x\dd{t'}/m_e c^2$; and the red dashed line denotes the work done by the $E_y$-component, $W_y(t) = -e\int_0^{t}E_y v_y\dd{t'}/m_e c^2$. In each of the bottom panels, the particle's trajectory is displayed by a rainbow-colored line with the progress of time shown in the colorbar (blue: earlier, red: later). The particle trajectories are overlaid on three simulation snapshots of $E_x$ to show the background field the particle is passing through.

For the VS case (velocity shear only, Fig.~\ref{fig:VS_track}), we select a particle that has reached a Lorentz factor of $10$ at $tc/L_x=27.4$ for display. As observed in the top panel, the Lorentz factor of this particle remains at $\gamma\sim 10$ throughout the simulation; it does not experience net acceleration over time. There is no substantial work done by the ideal motional $E$-field. The work done by $E_x$ and $E_y$ is anti-correlated and exhibits short and long periodicity on the timescales of $\sim L_x/c$ (short) and $10 L_x/c$ (long). In the bottom panels, which trace the trajectory of the particle at three time windows, we observe that the particle is initially undergoing $E\cross B$ drift (bottom left panel). When KH vortices develop, the background $E$-field changes, bending the drift's trajectory (bottom middle panel). The particle goes back and forth between two vortices with different field polarities in ellipsoidal motion, gaining and losing the same amount of energy as it passes through them. This leads to the short periodicity in the $E_x$ and $E_y$ work done. The particle maintains its ellipsoidal motion throughout the rest of the simulation, gaining no net energy. Beginning at $tc/L_x\approx 20$, some bigger bumps appear in the $E_x, E_y$ work done, with longer periodicity. This is due to waves passing by, as illustrated in the top right panel, which lead to neither acceleration nor any substantial change in the particle's trajectory. Therefore, we did not investigate these waves further.

For the MS case (magnetic shear only, Fig.~\ref{fig:MS_track}), we select a particle that has reached a Lorentz factor of $100$ at $tc/L_x=27.4$ for display, corresponding to the nonthermal tail in Fig.~\ref{fig:power_law}. As observed in the top panel, there is substantial increase in the Lorentz factor of this particle from $25$ to~$100$. The increase in $\gamma$ occurs in a single burst at $2.5\leq tc/L_x\leq 4.5$, mainly due to the work done by the $E_x$ component of the ideal motional $E$-field. There is hardly any work done by the $E_y$ component. In the bottom panels, we observe that the particle is initially performing a Speiser-like orbit close to the shear layer at $200 d_{e,j}$ due to the $B_z$-field reversal. When DK instability develops, the DK plumes bring plasma from both sides of the interface with the same $E_x$ polarity into alignment, creating a channel that accelerates the particles linearly \citep{Zenitani_Hoshino-2007}. In the saturated stage, a thickened shear layer is created. The EM field within the thickened layer is drastically reduced, so particles simply pass through it ballistically. When the particle exits the shear layer, it resumes its Speiser-like orbit under the influence of the $B$-field.

For the mixed shear case ($u_j=0.3, B_w/B_j=-1$, Fig.~\ref{fig:uj0.3_track}), we again select a particle that has reached a Lorentz factor of $100$ at $tc/L_x=27.4$ for display, corresponding to the nonthermal tail in Fig.~\ref{fig:power_law}. As observed in the top panel, there is substantial increase in the Lorentz factor of this particle from $35$ to $\sim 120$. The increase in $\gamma$ occurs in a stochastic manner, accumulating energy over an extended period of time. Most of the energy is gained over $12\leq tc/L_x\leq 25$, mainly due to work done by the $E_y$ component. This $E_y$ arises from the motional field $(-\vb{v}\cross\vb{B})$, which is non-zero due to the background fluid flow $v_x\vu{x}$ and magnetic field~$B_z\vu{z}$. At the same time, $E_x$ contributes very little to the overall increase in energy. In the bottom panels, we observe that the particle is initially performing a Speiser-like orbit (with a rather large radius of curvature), gradually shifting to an S-shaped trajectory in the nonlinear stage. Due to the moderately strong velocity shear in the background, the DK plumes are unable to align plasma with the same $E_x$ polarity to produce a large-scale $E_x$ channel, thus we do not observe linear acceleration by $E_x$ in the manner observed in the MS case. The background flow is substantially more turbulent than the MS and VS cases, and the $E_x$ field is also stochastic in appearance. In the top right panel, we display the initial positions of the particles that underwent substantial acceleration, defined by acquiring a Lorentz factor above~30 by $tc/L_x=27.4$, i.e., much higher than the mean $\bar{\gamma} \simeq 2.4$ corresponding to a thermal distribution with temperature $\theta=1$. These particles are clustered initially at the shear interface, implying that they are likely accelerated there.

\begin{figure}
    \centering
    \includegraphics[width=\textwidth]{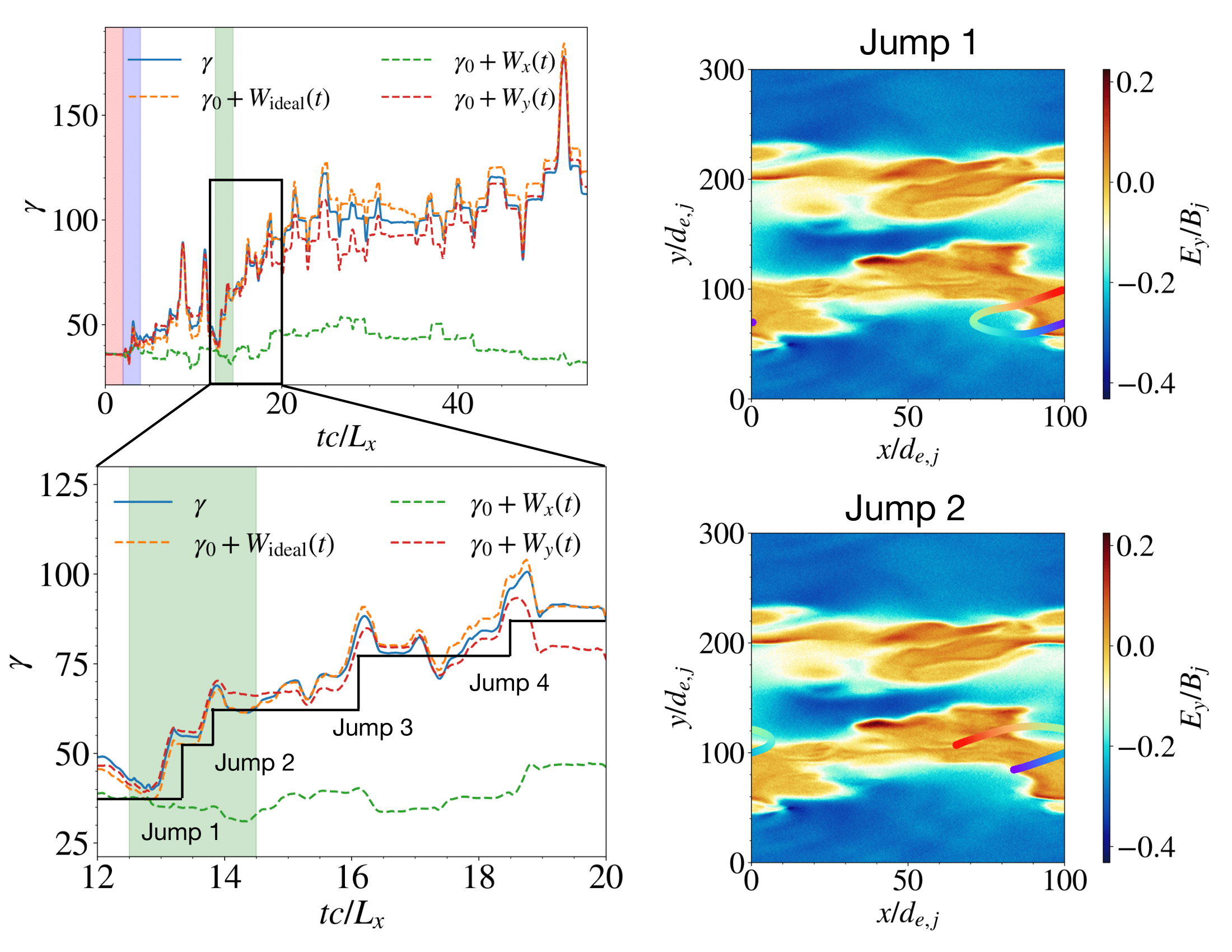} \\
    \includegraphics[width=0.9\textwidth]{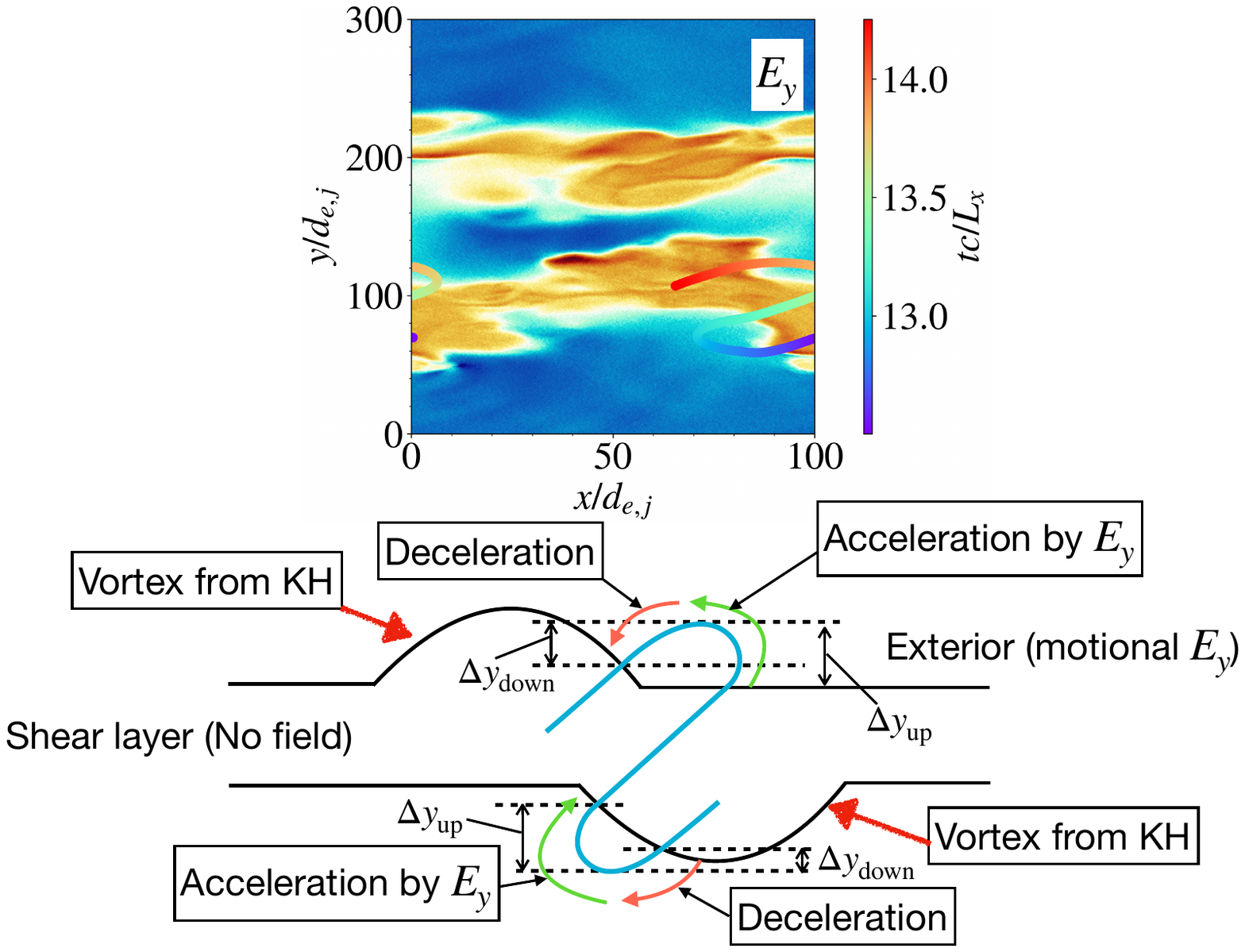}
    \caption{Top panels: (Left) Zoom-in energetics of the selected particle over $12\leq tc/L_x\leq 20$, illustrating the four jumps it took to acquire an increase in the Lorentz factor of $\sim50$. (Right) Particle trajectories over Jump 1 and 2, overlaid on two $E_y$ snapshots, with the colorbar indicating the strength of $E_y$. Bottom panels: Particle trajectory over Jump 1 and 2 combined, with a schematic diagram showing how it is energized throughout the process.}
    \label{fig:uj0.3_traj}
\end{figure}

Figs.~\ref{fig:VS_track}-\ref{fig:uj0.3_track} show that dissipation and particle acceleration in the single shear cases (VS and MS, where there is only one kind of shear) are very different from the mixed shear case. Specifically, in the mixed-shear case we uncovered a new mechanism by which particles are accelerated mainly due to work done by~$E_y$. This process is stochastic, with particle trajectories that exhibit a distinct S-shape. We illustrate the details of this type of acceleration in Fig.~\ref{fig:uj0.3_traj} by zooming in to the time period $12\leq tc/L_x\leq 20$, where the particle acquires the greatest increase in~$\gamma$. We observe an increase in $\gamma$ from $\approx 40$ to $\approx 90$, acquired over four major jumps. Each jump is characterized by net positive work done by~$E_y$. Focusing on jump 1 and jump 2, in the two top right panels (titled `Jump~1' and `Jump~2') of Fig.~\ref{fig:uj0.3_traj} we plot the trajectories of the particle as it is energized, overlaid on the $E_y$ field background, with the colorbar indicating the strength of $E_y$. The trajectories are again rainbow-colored to indicate the progress of time (blue: early, red: later). The thickened shear layer is clearly marked in orange in this color scheme, and we observe $E_y\approx0$ within the layer, so particles are not energized or de-energized by $E_y$ as they pass through the layer. A finite $E_y<0$ is present outside the shear layer, marked by blue, mainly due to the motional $E$-field. During jump 1, the particle performs a U-shape turn, exiting the shear layer in the first half of the displayed trajectory and reentering it in the second half. As the tracked particle is an electron and $E_y<0$, there is a slight de-energization as the particle exits the shear layer, going in the $-y$ direction. The electron quickly turns around due to~$B_z$, going in the $+y$ direction, and is then energized before reentering the shear layer. Due to the bent geometry of the shear layer, the particle covers a greater distance going upward than downward in the region with finite $E_y$ (blue region), thus acquiring net energy. The same thing happens at jump 2, where the particle traverses a larger distance going upward than downward in the region with finite $E_y$ due to the bent geometry of the shear layer, thus acquiring net energy. This asymmetry between energization and de-energization of the particle in the motional $E_y$ field, due to the bent geometry of the shear layer, leads to a gradual accumulation of energy and, thus, to particle acceleration. Combining the particle trajectories through jumps 1 and~2, we observe an S-shaped pattern, with the particle weaving in and out of the thickened shear layer. 

We offer a simplified model of how particles are accelerated in the mixed-shear cases with the help of the schematic diagram at the bottom of Fig.~\ref{fig:uj0.3_traj}. As shown in the diagram, a particle performs an S-shaped trajectory in and out of the shear layer due to the presence of a $B_z$ field outside the shear. The particle streams through the shear layer ballistically as there is no field within it. Due to the bent geometry of the shear layer (depicted by the two lumps above and below the shear layer in the diagram), there is asymmetric energization and de-energization as work done is given by $\int \vb{E}_y\cdot\dd{\vb{l}}\approx E_y(\Delta y_\mathrm{up} - \Delta y_\mathrm{down})$, where $\Delta y_\mathrm{up}$ and $\Delta y_\mathrm{down}$ are the upward and downward displacements of the particle as it weaves in and out of the bent shear layer (Fig.~\ref{fig:uj0.3_traj}).  If $\Delta y_\mathrm{down}<\Delta y_\mathrm{up}$ there will be a net energization. On the other hand, if the shear layer were not bent (i.e. flat in the $x$-direction, as in the saturated stage of the simulation, $tc/L_x\gtrsim 20$), then there would be equal energization and de-energization of the particle as it weaves in and out of it, resulting in no particle acceleration. The DK and KH instabilities both contribute to creating such a bent layer; in particular, DKI creates a shear layer where the electromagnetic field is suppressed, while KHI creates the vortical modulations which gives the layer a bent geometry; without either of them we would not have observed such particle acceleration. In Appendix \ref{app:additional_example} we show additional examples of particle acceleration through this mechanism, illustrating that the process described is not an isolated event.

\begin{figure}
    \centering
    \includegraphics[width=0.45\textwidth]{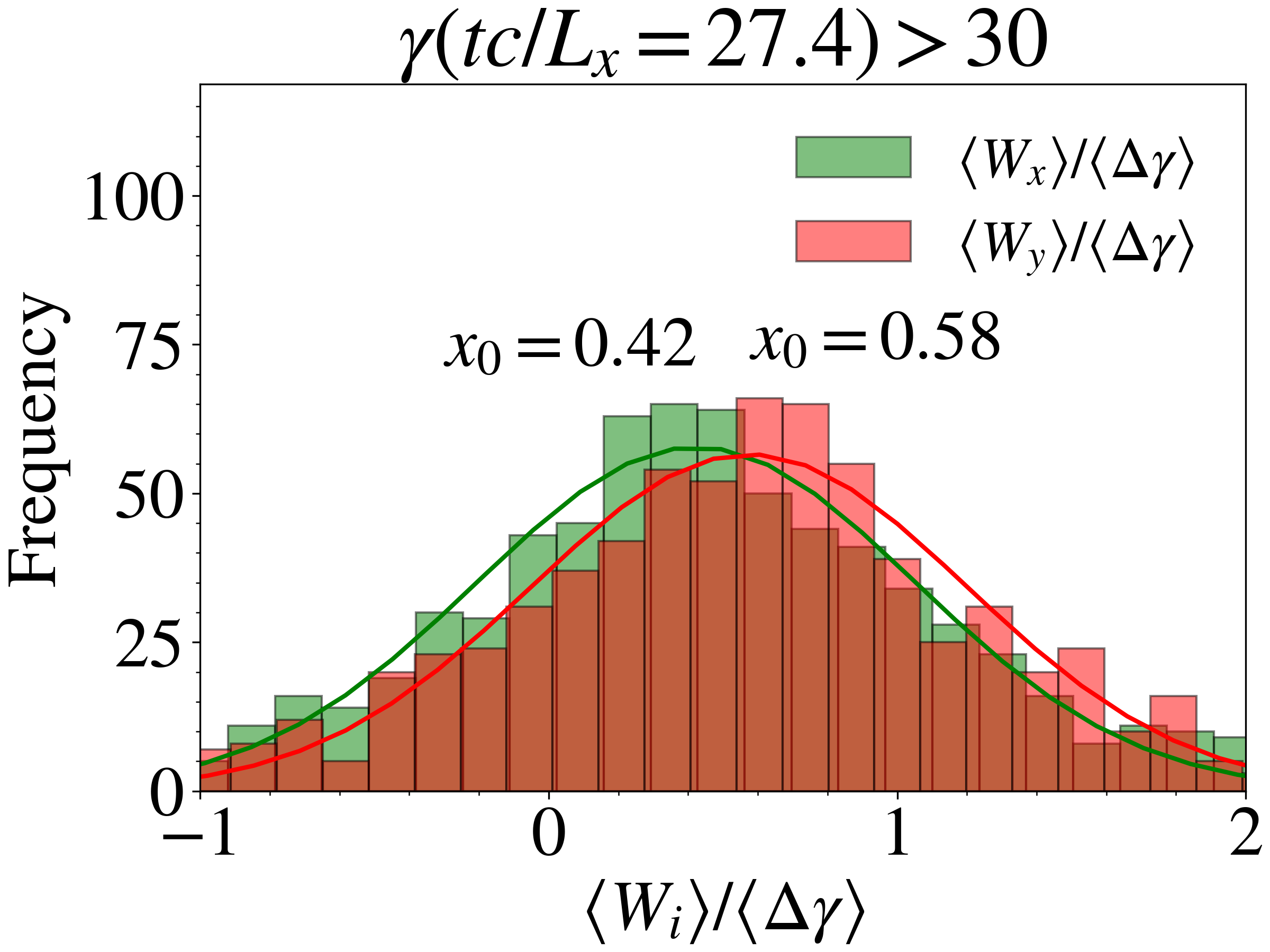}
    \includegraphics[width=0.45\textwidth]{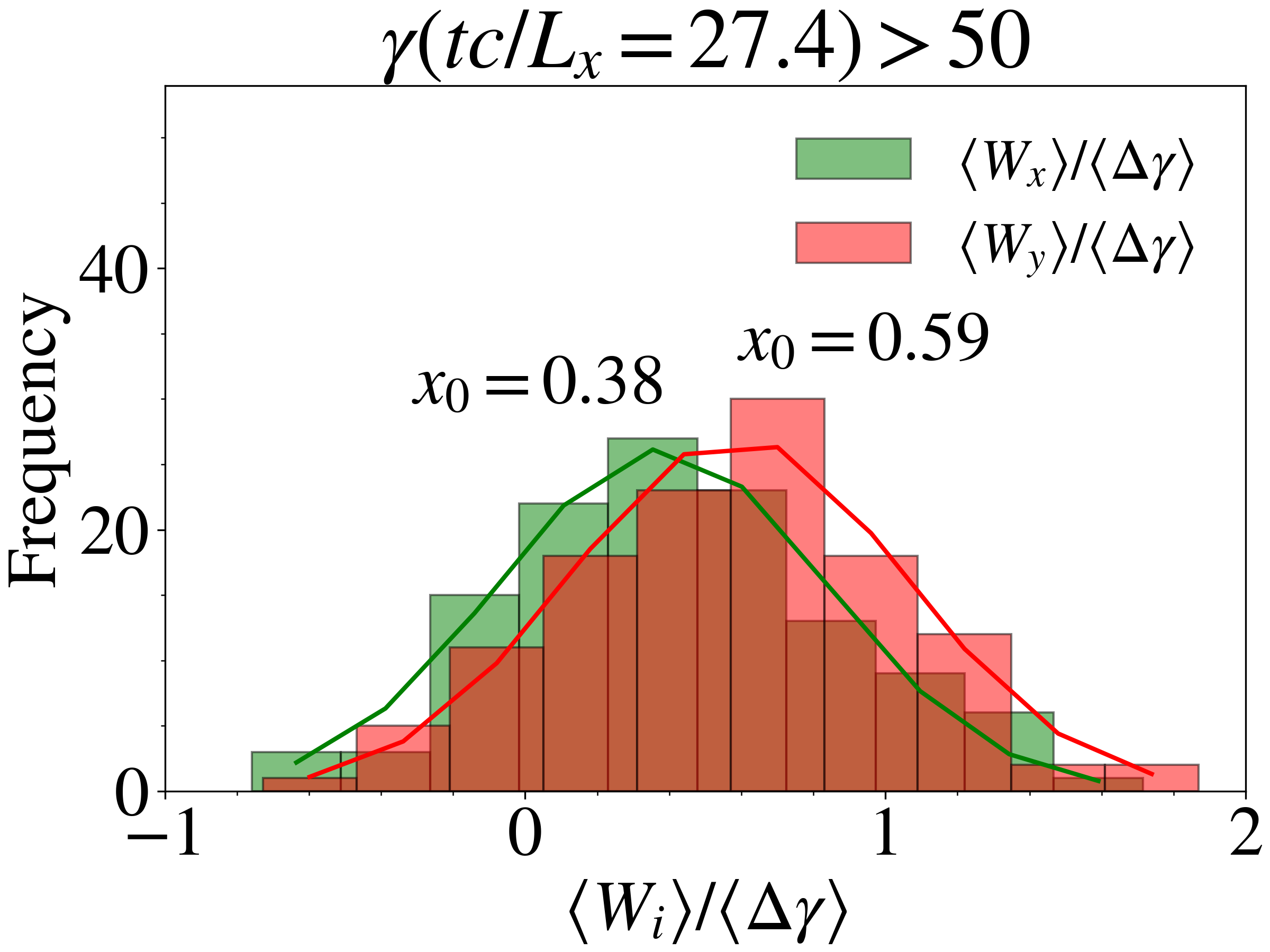} \\
    \includegraphics[width=0.45\textwidth]{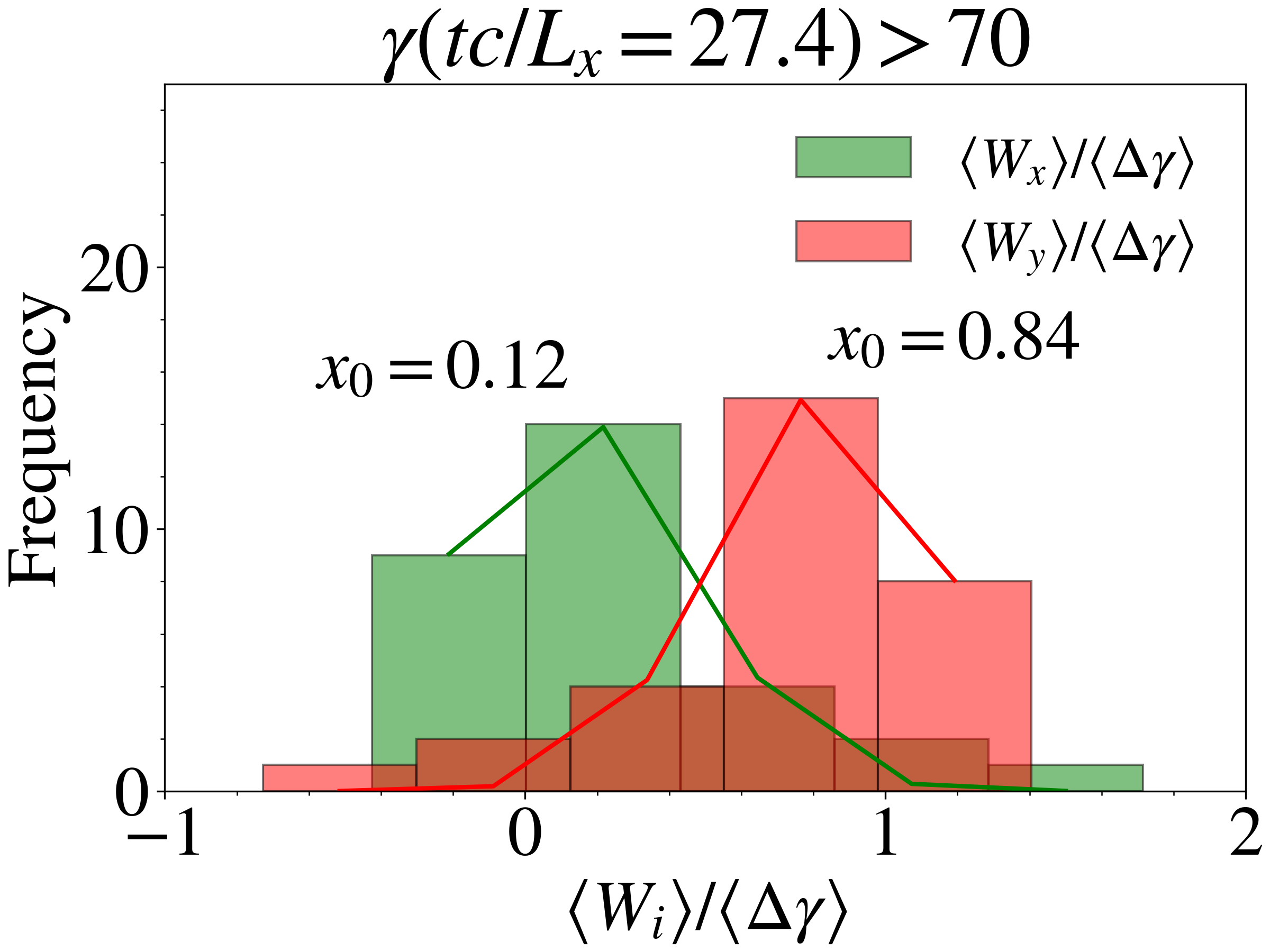}
    \caption{Histograms showing contributions of the work done by $E_x$ (green) and $E_y$ (red), denoted respectively by $W_x, W_y$ (defined in \S\ref{subsec:ptcl_acc}), to the increase in the Lorentz factor $\gamma$ ($\Delta\gamma=\gamma - \gamma(t=0)$), for tracked particles that have acquired $\gamma>30$ (top left), $\gamma>50$ (top right), and $\gamma>70$ (bottom) at $tc/L_x=27.4$. The histograms are fitted with a Gaussian $\exp(-(x - x_0)^2/2\sigma^2)$ and the fitted shifts $x_0$ are displayed. The angled brackets $\langle\cdot\rangle$ denote time-average over $30\leq tc/L_x\leq 50$.}
    \label{fig:hist_contribution}
\end{figure}

The mechanism described, in which particles are accelerated due to asymmetric work done by the motional $E_y$ external to the bent shear layer as they traverse an S-shaped orbit through it, is not the only way particles are accelerated, but accounts for particles at the highest end of the nonthermal spectrum. The histograms in Fig.~\ref{fig:hist_contribution} show contributions of the work done by $E_x$ (green) and $E_y$ (red), denoted by $W_x, W_y$, to the increase in the Lorentz factor $\Delta\gamma$, for tracked particles that have acquired $\gamma > 30$ (top left), $\gamma > 50$ (top right), and $\gamma > 70$ (bottom) at $tc/L_x=27.4$. The angled brackets $\langle\cdot\rangle$ denote time-average over $30\leq tc/L_x\leq 50$. Thus, if a particle has value of $\langle W_y\rangle/\langle\Delta\gamma\rangle \approx 1$, it means that all of the increase in $\gamma$ is due to work done by $E_y$, and vice versa when this value is zero. Note that the ratio $\langle W_i\rangle/\langle\Delta\gamma\rangle$ can be greater than 1 or less than 0 as there could be net negative or positive work done by the other component of the electric field. We fit the histograms with a Gaussian $\exp[-(x-x_0)^2/2\sigma^2]$ and the fitted shifts $x_0$ are displayed. Observe that in all three panels, $W_y$ contributes the majority to the increase in particle energy. The discrepancy between $W_y$ and~$W_x$, as shown by the difference in~$x_0$, increases as we move to higher energy thresholds (from $\gamma>30$ to $\gamma>70$). In particular, for particles with $\gamma>70$, $W_y$ contributes $84\%$ of the energy gained while $W_x$ contributes only~$12\%$. Thus, the mechanism described above accounts for most of the particle acceleration in the mixed-shear cases, particularly for the highest energy particles.

\begin{figure}
    \centering
    \includegraphics[width=0.6\textwidth]{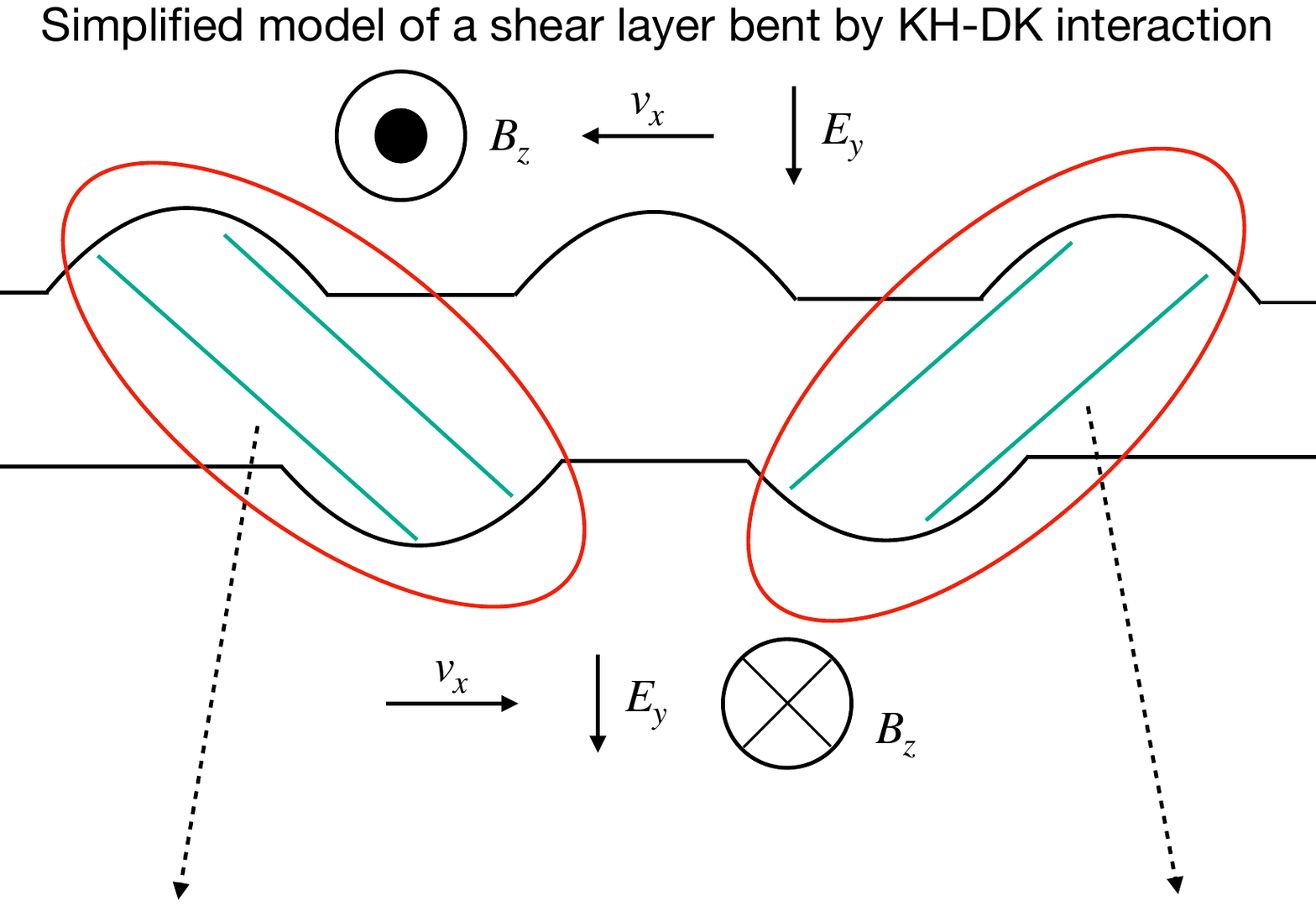} \\
    \includegraphics[width=0.45\textwidth]{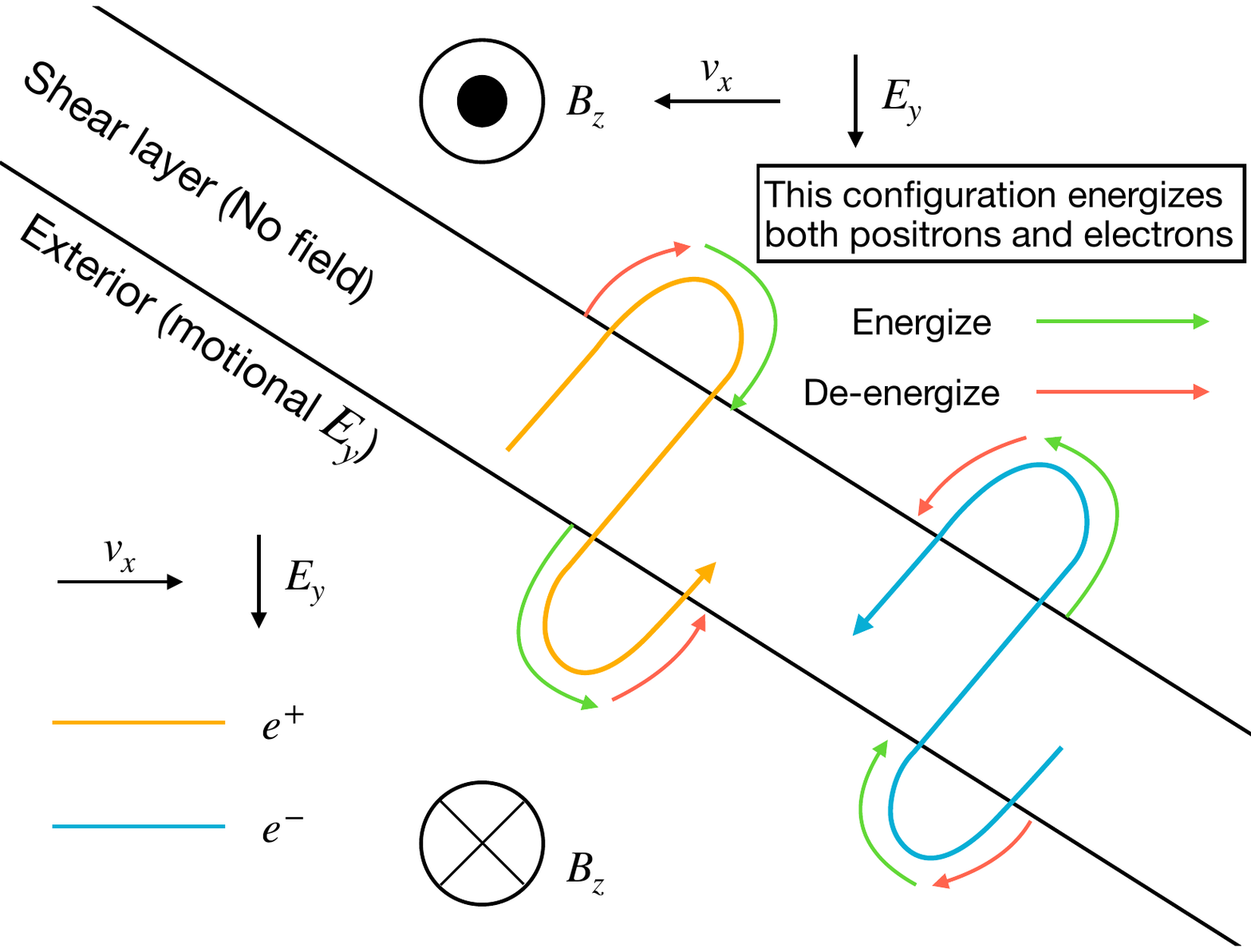}
    \includegraphics[width=0.45\textwidth]{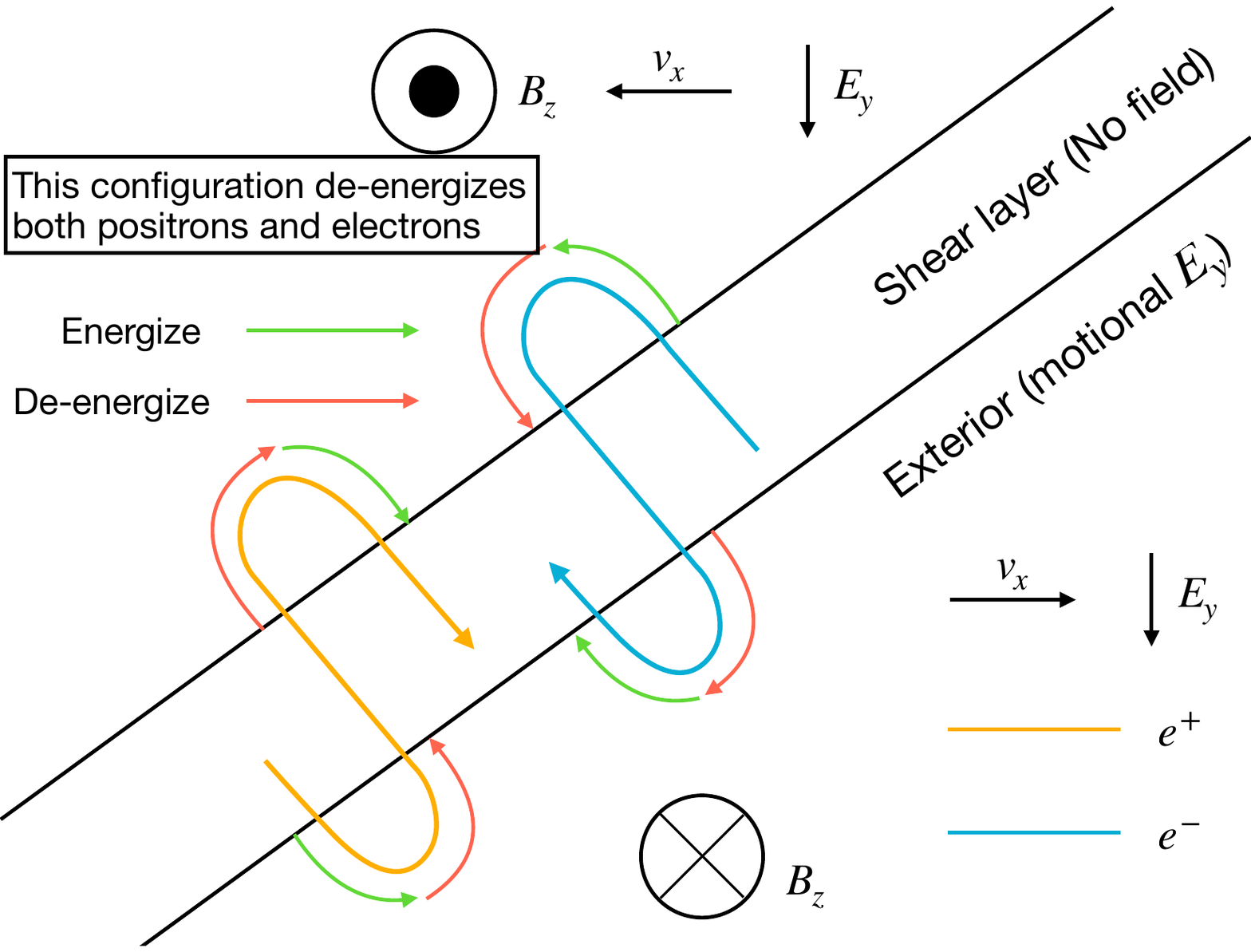} \\
    \includegraphics[width=0.6\textwidth]{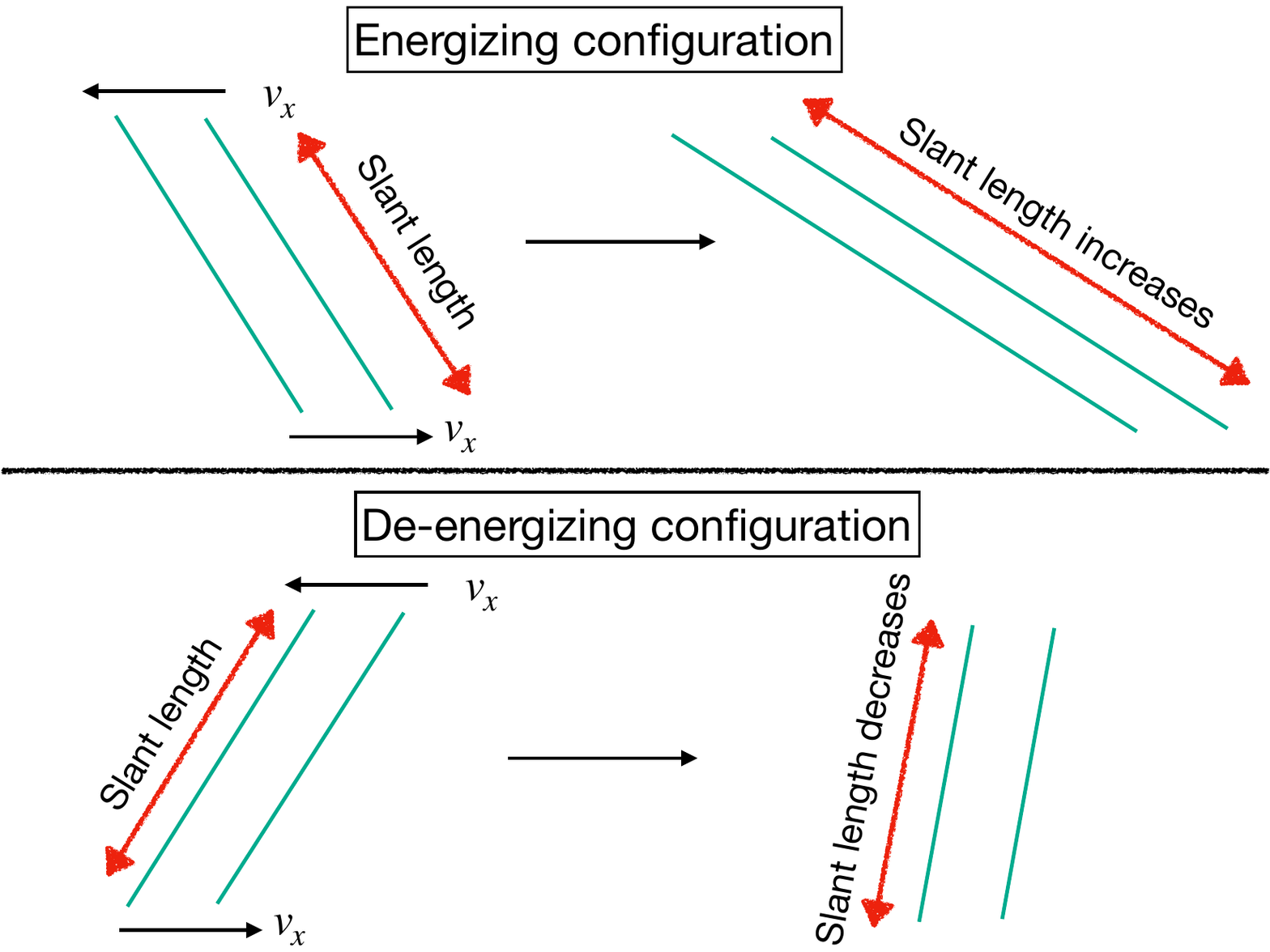}
    \caption{Top panel: Simplified model of a shear layer bent by KH-DK interaction, where regions of possible acceleration/deceleration (circled in red) are modeled as slanted strips. Middle panels: Depending on the orientation of the slanted strips (left: downwards from left to right; right: upwards from left to right), electrons and positrons can be net energized (left) or de-energized (right) due to asymmetric work done by the motional $E_y$ field exterior to the shear layer. Bottom panel: The slant length of the strip in the energizing configuration is stretched by background shear while it is shrunk in the de-energizing configuration.}
    \label{fig:configurations}
\end{figure}

The configuration depicted in the bottom panel of Fig.~\ref{fig:uj0.3_traj}, where the particle exits and re-enters the shear layer with the slant direction as shown, favors net energization of the particle. If the slant direction of the shear layer is different, net de-energization may result. In Fig.~\ref{fig:configurations} we describe when this happens. Focusing on a simplified picture where regions of possible acceleration/deceleration in a shear layer bent by the KH-DK interaction are modeled as diagonal strips (see top panel), we observe that the configuration in the middle left panel, where the shear layer slants down from left to right, \emph{energizes} both electrons and positrons, while the other configuration, where the shear layer slants up from left to right, \emph{de-energizes} them. One can convince oneself that the direction the particle is traveling initially does not affect whether it is energized or de-energized--it is determined only by the slant direction of the shear layer. Thus, if particles (both electrons and positrons) encounter the configuration on the left more often than the right one, net stochastic acceleration results.  Intuitively, the slant length of the shear layer in the energizing configuration will be stretched out due to the background shear, while in the de-energizing configuration it will shrink (see bottom panel of the figure). A larger slant length will likely expose the particles to more energizing encounters, thus resulting in net energization. We conclude this section with a remark that if one flips the orientation of the magnetic field while keeping the velocity field the same, the exterior motional electric field will flip in sign but the trajectories of the electrons and positions will reverse such that overall the energizing configuration will remain energizing and vice versa for the de-energizing configuration.

\section{Discussion and conclusion} \label{sec:conclusion}

In this study, we performed 2D relativistic, collisionless pair-plasma PIC simulations of a thin layer subject to velocity and magnetic shear in the relativistically warm ($\theta= T/m_ec^2\sim 1$) and moderately magnetized ($\sigma\sim 1$) regime. We created conditions where only the Kelvin-Helmholtz (KH) and Drift-Kink (DK) instabilities can develop, while tearing modes are prohibited. By varying the magnitude of the velocity shear from zero to some supersonic value while keeping the magnetic field equal but opposite across the shear layer, we investigated the effects of velocity shear on~DKI, including its nonlinear interplay with~KHI. Our main findings are:

\begin{itemize}
    \item The interaction of KH and DK instabilities creates qualitatively new flow structures compared to the cases that harbor only KHI or DKI: a thick, pressure-dominated shear layer with very weak electromagnetic field (`annihilated core'), modulated by KH vortices (`KH cocoon').
    \item A moderate velocity shear ($u_j\leq 0.4$) has minimal effect on the linear growth of DK modes due to the lag between DK and KH growth times. For $u_j>0.4$, the linear evolution is substantially modified, and no exponential growth phase can be captured.
    \item The dissipation of magnetic energy and bulk kinetic energy varies drastically with velocity shear and depends sensitively on how strongly the KH and DK instabilities are excited. In the absence of velocity shear (MS case), DKI can dissipate a significant amount of magnetic energy. However, a weak shear ($u_j\leq 0.1$) disrupts DKI plumes and thus reduces dissipation. For moderately strong velocity shear ($0.2\leq u_j\leq 0.4$), KHI becomes excited and KH vortices lead to a drastic rise in dissipation. Finally, for supersonic shear, KHI is weakened and we observe reduced dissipation. In the most dissipative case ($u_j=0.3$), 40\% of the magnetic energy and 50\% of the bulk kinetic energy within the box are converted to internal energy while these numbers can be as low as 10\% for some other velocity shear values. We note that less than 5\% of the magnetic energy and nearly zero bulk-kinetic energy are dissipated in the VS (purely KH) case. More details can be found in \S\ref{subsec:morp_diss} (Fig.~\ref{fig:schematic}).
    \item We observe nonthermal power-law tails in the energy spectra of the mixed-shear cases, distinct from the MS (purely DK) and VS (purely KH) cases, thus suggesting different particle-acceleration mechanisms. By examining the trajectories and energization histories of selected particles that have been accelerated, we find that in the MS case particles are accelerated linearly in a channel with aligned $E_x$ (consistent with \citet{Zenitani_Hoshino-2007}) while in the mixed-shear case particles are stochastically accelerated by $E_y$ in a shear layer bent by KH-DK interaction. This is a newly discovered way to accelerate particles.
    \item In this new acceleration mechanism, particles weave in and out of a bent shear layer in an S-shaped pattern due to the exterior magnetic field. They gain energy when the re-entry location is downstream/upstream (for positrons/electrons) of the exit location (relative to the exterior motional electric field) and lose energy in the opposite case. Whether a particle gains or loses energy is \emph{independent} of the direction it is traveling---it depends only on the slant direction of the shear layer (as depicted in Fig.~\ref{fig:configurations}). The slant configuration that energizes particles is more likely to be found than the opposite, thus leading to net acceleration.
    \item Other acceleration mechanisms exist in the mixed-shear cases, but the new mechanism accounts for the majority of the particles accelerated and for the highest energy particles found.
\end{itemize}

How are our results likely to scale with system size? Focusing on the most dissipative mixed-shear case ($u_j=0.3$), we note that there is a high-energy cut-off in the particle energy spectrum at $\gamma\sim300$. Is this cut-off intrinsic to the acceleration mechanism or due to limited system size? Note that for the assumed conditions in zone j ($\theta_j=1,\sigma_j=1$), $d_{e,j}\sim\rho_{0,j}$, where $\rho_{0,j}=m_e c^2/eB_j$ is the nominal gyroradius. For relatively unchanged magnetic field exterior to the shear layer, the system size can be cast as $L/\rho_{0j}\sim L/d_{e,j}\sim 100$. The Hillas energy limit, which is defined as $\gamma_\mathrm{max}=L/\rho_{0j}$, is then $\sim 100$, comparable to the observed spectral cutoff. 
The high energy cut-off is thus likely due to the system-size limit, which suggests that the nonthermal power-law tail in the mixed-shear case will likely extend further with a larger box size. It follows that in astrophysical settings particles can be accelerated to well beyond $\gamma\sim 10^2$. We relegate a detailed examination of system size scaling to future work.

In this study we explored only a relativistically warm, moderately magnetized pair-plasma in 2D, varying the velocity shear while holding magnetic shear constant. We also restricted our consideration to KH and DK instabilities, leaving out tearing modes. Many possible follow-up studies can be devised, for example by varying the initial orientation of the magnetic field ($\vb{B}=B_z\vu{z}\rightarrow \vb{B} = B(\cos\theta\vu{x} + \sin\theta\vu{z})$) so as to introduce tearing modes into the problem. One can also investigate how the results translate to a normal ($e-p$) plasma and across a range of temperatures and magnetizations. Ultimately, the aim is to generalize our incremental understanding of the problem from 2D to~3D. This study, therefore, should be viewed as only the first step towards a comprehensive understanding of dissipation and particle acceleration at intermittent structures which harbor both velocity and magnetic shears.

The shear layers with which we initiate our setup are marked by spatially coinciding vortex and current sheets with the same thickness. While this is a simplifying assumption, \citet{Jain_etal-2021,Hubbert_etal-2021} have shown that kinetic-scale current sheets are tightly correlated to electron shear flow structures, at least in non-relativistic, $e-p$ plasma turbulence. In fact, thin vortex and current sheets have separately been found to be characteristics of intermittency in turbulence, and responsible for a substantial amount of dissipation \citep{Zhdankin_etal-2016}. A detailed examination of the alignment and magnitudes of current and vortex sheets at intermittent structures, in simulations of MHD turbulence, would offer us a clearer picture  the most relevant configurations for future simulations. 

Our results are relevant to a range of astrophysical settings because of the ubiquity of plasma turbulence. In low-energy settings, dissipation at intermittent structures has often been associated with nanoflares, X-ray bursts in the solar corona and wind \citep{Dmitruk_Gomez-1997,Dmitruk_etal-1998}. Recent studies have begun to see intermittent structures as the means by which cosmic rays in galaxies are scattered and transported \citep{Kempski_etal-2023,Lemoine_2023}. In high-energy settings, intermittent heating could be responsible for flares observed in black-hole accretion flows \citep{Giannios-2012, Ripperda_etal-2020}, limb brightening at AGN jet boundaries \citep{Duran_etal-2016,Sridhar_etal-2024}, rapid variability in gamma-ray bursts \citep{Drenkhahn_Spruit-2002}, and other phenomena, thus highlighting the need for further investigation of the nonlinear interactions of multiple instabilities, sourced from magnetic and velocity shears, at the kinetic level.

\emph{Acknowledgements.}--We thank Lorenzo Sironi, Vladimir Zhdankin and Navin Sridhar for useful comments and suggestions. This work is supported by NASA Astrophysics Theory Program  Grants 80NSSC22K0828 and 80NSSC24K0941, and ACCESS computing grants PHY140041 and PHY240194.

\appendix 

\section{Details of the model setup} \label{app:setup}

Our setup is based on the one described in~\cite{Rowan-2019}. 
Due to velocity and magnetic shears, there are current and charge excesses at the shear interfaces, which imply that the electron and positron densities and bulk velocities will be different close to the interfaces. Therefore, when we mention quantities with the subscripts `j' or `w' we are referring to their values far from the interfaces, in the respective zone `j' and `w' regions where the current and charge excess are zero. There, the bulk speeds and rest-frame electron and positron densities are equal: $\beta_{e,j}=\beta_{p,j}$, $\beta_{e,w}=\beta_{p,w}$ and $\tilde{n}_{e,j} = \tilde{n}_{p,j}$, $\tilde{n}_{e,w} = \tilde{n}_{p,w}$. 

Assuming equal jet rest-frame electron and positron temperatures, $\theta_{e,j} = \theta_{p,j}\equiv\theta_j$, 
the rest-frame mean particle speed $\bar{v}_j$ in Zone j's (and the mean Lorentz factor~$\bar{\gamma}_j$) for a given jet temperature $\theta_j$ can be determined from the local Maxwell-J\"uttner distribution. The adiabatic index $\Gamma_\mathrm{ad}(\theta_j)$
of the electrons and positrons can be found using the fitting formula given by eq.~14 in \cite{Service-1986}. 
Together,  $\bar{\gamma}_j$ and the rest-frame jet electron (and positron) particle density $\tilde{n}_{e,j} = \tilde{n}_{p,j} = \tilde{n}_{e,j}$ define the main normalizing length-scale in our simulations: the electron inertial length $d_{e,j} \equiv c/\omega_{pe,j}$, where $\omega_{pe,j} \equiv (4\pi\tilde{n}_{e,j}e^2/\bar{\gamma}_j m_e)^{1/2}$ is the comoving electron plasma frequency.
The comoving jet enthalpy density, which enters the expression for magnetization~$\sigma_j$, is given by $w_j = w_{e,j} + w_{p,j} = 2(1+\Gamma_{\mathrm{ad},j}/(\Gamma_{\mathrm{ad},j}-1)\theta_j)\tilde{n}_{e,j}m_e c^2$.  

The plasma bulk velocity $\beta_x\qty(y)$ and the magnetic field are initialized with the following profiles:
\begin{gather}
    \beta_x\qty(y) = \beta_j\qty[1 - \tanh(\frac{y-y_1}{\Delta}) + \tanh(\frac{y-y_2}{\Delta})], \\
    B_z\qty(y) = B_j\biggl\{\frac{B_w}{B_j} + \frac{1}{2}\qty(1 - \frac{B_w}{B_j})\biggl[\tanh(\frac{y-y_1}{\Delta}) \nonumber\\ - \tanh(\frac{y-y_2}{\Delta})\biggr]\biggr\},
\end{gather}
where $\beta_j = u_j/(1 + u_j^2)^{1/2}$ is the velocity of Zone j's (normalized by~$c$) corresponding to the specified normalized 4-velocity~$u_j$; $y_1,y_2$ are the $y$-locations of the shear interfaces; and $\Delta$ is the half-width of the shear (and also current) layer at these interfaces. 
%
The initial lab-frame electric field, current density, and charge density are given by $\vb{E}=-\vb{v}\cross\vb{B}/c= \beta_x B_z\vu{y}$, $\vb{J} = c\dv*{B_z}{y}/4\pi\vu{x}$, and $\rho_e = (\dv*{E_y}{y})/4\pi$, as dictated by the ideal-MHD Ohm's law, Ampere's law, and Gauss's law, respectively. 

The $y$-profiles of the lab-frame electron and positron number densities $n_e, n_p$ and of the bulk velocities $\beta_{e,x}, \beta_{p,x}$ are constrained to satisfy the self-consistency conditions
\begin{gather}
    \rho_e = \qty(n_p - n_e) e, \label{eqn:rho_e}\\
    J_x = \qty(n_p \beta_{p,x} c - n_e \beta_{e, x}) e c, \\
    \beta_x = \frac{n_p\beta_{p,x} + n_e\beta_{e,x}}{n_p + n_e}. \label{eqn:mean_vel}
\end{gather}
To proceed, we recast $n_p, n_e$ as
\begin{equation}
    n_p = n_0\qty(1-\delta_n), \quad n_e = n_0\qty(1+\delta_n), \label{eqn:density_recast}
\end{equation}
where $n_0 = \Gamma(y)\tilde{n}_0$ can be regarded as some background lab-frame density profile. The bulk Lorentz-factor profile is given by $\Gamma(y) = 1/(1 - \beta_x^2)^{1/2}$, and the rest-frame background density $\tilde{n}_0$ by
\begin{gather}
    \tilde{n}_0\qty(y) = \tilde{n}_{e,j}\biggl\{\frac{\tilde{n}_{e,w}}{\tilde{n}_{e,j}} + \frac{1}{2}\qty(1 - \frac{\tilde{n}_{e,w}}{\tilde{n}_{e,j}})\biggl[\tanh(\frac{y-y_1}{\Delta}) \nonumber\\ - \tanh(\frac{y-y_2}{\Delta})\biggr]\biggr\}.
\end{gather}
In this study, for simplicity we take the rest-frame jet and wind density to be the same, $\tilde{n}_{e,w}/\tilde{n}_{e,j} = 1$. Substituting eq.~\ref{eqn:density_recast} into eqs.~\ref{eqn:rho_e}-\ref{eqn:mean_vel} and solving, we then have
\begin{gather}
    \delta_n = -\frac{\rho_e}{2n_0 e}, \\
    \beta_{e,x} = \frac{2n_0 e\beta_x - J_x/c}{2n_0 (1 + \delta_n)e}, \quad \beta_{p,x} = \frac{2n_0 e\beta_x + J_x/c}{2n_0 (1 - \delta_n)e}.
\end{gather}
Finally, to determine the electron and positron temperature profiles $\theta_{e,p}(y)$, we employ pressure balance:
\begin{equation}
    \tilde{n}_p m_e c^2\theta_p + \tilde{n}_e m_e c^2\theta_e + \frac{B_z^2}{8\pi\Gamma^2} = \mathrm{const.} \label{eqn:pressure_balance}
\end{equation}
Assuming $\theta_p(y) = \theta_e(y)$ initially, these temperatures can be determined from eq.~\ref{eqn:pressure_balance}. 
This completes the PIC setup of the model.

\section{Definition of the plasma microscales} \label{app:microscales}

The key plasma length-scales describing our system are the initial jet electron inertial length 
$d_{e,j} \equiv c/\omega_{pe,j}$, where $\omega_{pe,j} \equiv (4\pi \tilde{n}_{ej} e^2/\bar{\gamma}_jm_e)^{1/2}$ 
is the plasma frequency ($\tilde{n}_{e,j}$ is the rest-frame jet electron number density and $\bar{\gamma}_j$ is the mean Lorentz factor of Zone j's rest-frame Maxwell-J\"uttner distribution), 
Zone j's electron Debye length $\lambda_{\mathrm{D}e} \equiv (T_j/4\pi n_{e,j} e^2)^{1/2}\sim\theta_j^{1/2} d_{e,j}$, 
and the characteristic jet electron gyroradius $\rho_j \equiv \bar{\gamma}_j m_e \bar{v}_j c/e B_j\sim (\theta_j/\sigma_j)^{1/2} d_{e,j}$ (where $\bar{v}_j$ is the mean particle velocity). For relativistically warm ($\theta\sim 1$), moderately magnetized ($\sigma\sim 1$) plasma the three plasma length-scales ($d_{e,j},\lambda_{\mathrm{D}e},\rho_j$) are roughly equal.

\section{Decomposing the stress tensor into thermal and bulk components} \label{app:decompose}

We describe the procedure for decomposing the stress tensor into thermal and bulk-flow components, needed to monitor the bulk-kinetic energy content. We thank Vladimir Zhdankin (private communication) for offering insights on this procedure.

The procedure is as follows:
\begin{enumerate}
    \item[1.] For each species $s$, calculate the number density $n = \sum_l w_l/\Delta V$, the stress tensor $\Pi_{ij} = \sum_l\gamma_l m_{s,l} w_l v_{l,i} v_{l,j}/\Delta V$, momentum density $U_{p,i} = \sum_l\gamma_l m_{s,l} w_l v_{l,i}/\Delta V$, energy density $U_e = \sum_l\gamma_l w_l m_{s,l} c^2/\Delta V$ and particle flux $F_i = \sum_l w_l v_{l,i}/\Delta V$ at each grid point, where the sum is taken over all (macro)particles in the neighborhood of a cell, $w_l$ is the weight of the macroparticles, $\gamma_l$ is their Lorentz factor, $m_{s,l}$ is the particle mass, $\Delta V$ is the volume element, and $v_{l,i}$ is the $i$-component of the particle's velocity. These lab-frame quantities are outputted automatically in \textsc{Zeltron}.
    \item[2.] Rotate the lab-frame vector and tensor quantities $\vb{U}_p$, $\vb{F}$, $\vb{\Pi}$ to the frame $x',y',z'$ such that the particle flux $\vb{F}'$ points in the $x'$-direction. Note that scalar quantities such as $U_e, n$ are rotationally invariant.
    \item[3.] The components of the comoving thermal pressure tensor in the rotated frame can be obtained by the following equations:
    \begin{align}
        P'_{x'x'} &= \Gamma_b^2\qty(\Pi'_{x'x'} - 2v_b U'_{p,x'} + \qty(v_b/c)^2 U_e), \nonumber\\
        P'_{x'y'} &= \Gamma_b\qty(\Pi'_{x'y'} - v_b U'_{p,y'}), \nonumber\\
        P'_{x'z'} &= \Gamma_b\qty(\Pi'_{x'z'} - v_b U'_{p,z'}), \nonumber\\
        P'_{y'y'} &= \Pi'_{y'y'}, \nonumber\\
        P'_{y'z'} &= \Pi'_{y'z'}, \nonumber\\
        P'_{z'z'} &= \Pi'_{z'z'}, \nonumber
    \end{align}
    where $v_b = F'_{x'}/n$ is the bulk speed of the Eckart frame (the frame for which particle flux is zero) and $\Gamma_b = (1 - v_b^2/c^2)^{-1/2}$.
    \item[4.] Inverse-rotate $P'_{i'j'}$ to obtain the thermal pressure tensor $P_{ij}$ in the lab frame. The bulk-flow dynamic pressure can be obtained by subtracting $P_{ij}$ from $\Pi_{ij}$ (i.e. $P_{b,ij} = \Pi_{ij} - P_{ij}$). These are then the thermal and dynamic pressure tensors for species~$s$.
    \item[5.] The total bulk-flow kinetic and thermal energies can then be obtained by taking the trace of the respective pressure tensors over all volume $\sum_V\sum_i P_{b,ii}\Delta V$ and $\sum_V\sum_i P_{ii}\Delta V$, for all species present.
\end{enumerate}

The justification for the procedure is as follows. First, note that thermal pressure is defined in the bulk frame of the plasma:
\begin{equation}
    P_{ij} = \int\dd^3\vb{\bar{p}} \bar{f} \bar{\gamma} m_s\bar{v}_i\bar{v}_j, \label{eqn:therm_press}
\end{equation}
with barred quantities denoting comoving-frame quantities. Here, $\bar{f}(\vb{\bar{x}},\vb{\bar{p}},\bar{t})$ is the distribution function satisfying $\bar{n} = \int\dd^3\vb{\bar{p}}\bar{f}$. The comoving frame is connected to the lab (or simulation) frame (denoted by unbarred quantities) by the Lorentz transformation corresponding to the bulk velocity $\vb{v}_b$ and the associated Lorentz factor $\Gamma_b = (1 - v_b^2/c^2)^{-1/2}$. We do not specify $\vb{v}_b$ for now to keep this derivation general. Using the Jacobian of the Lorentz transformation matrix and the fact that particle number is invariant with respect to frame transformation, we have $\dd^3\vb{p}/\gamma=\dd^3\vb{\bar{p}}/\bar{\gamma}$ and $f = \bar{f}$. This gives
\begin{equation}
    \int\dd^3\vb{\bar{p}} \bar{f} \bar{\gamma} m_s\bar{v}_i\bar{v}_j = \int\dd^3\vb{p}\frac{\bar{\gamma}}{\gamma} f \bar{\gamma} m_s \bar{v}_i\bar{v}_j = \int\dd^3\vb{p}\frac{f}{\gamma m_s}\bar{p}_i\bar{p}_j. \label{eqn:therm_pres_transform}
\end{equation}
If we assume the bulk motion is entirely in the $x$-direction, $\vb{v}_b=v_b\vu{x}$ (if it were not, one could simply rotate into a frame in which it is), then Lorentz transformation of the 4-momentum gives $\bar{p}_x=\Gamma_b(p_x - v_bE/c^2)$, $\bar{p}_y=p_y,\bar{p}_z=p_z$. Substituting into eq.~\ref{eqn:therm_pres_transform} and noting that $\Pi_{ij} = \int\dd^3\vb{p} f \gamma m_s v_i v_j$, $U_{p,i} = \int\dd^3\vb{p} f \gamma m_s v_i$, $U_e = \int\dd^3\vb{p} f \gamma m_s c^2$ in the continuous limit gives the equations listed in point 3 of the procedure.
Unlike the prescription described in \citet{Zhdankin-2021}, we have selected the bulk frame to be the zero-particle-flux frame (Eckart's frame), $v_b = F_x/n$.

\section{Additional examples of particle acceleration by the new mechanism in the mixed-shear $u_j=0.3$ case} \label{app:additional_example}

\begin{figure}
    \centering
    \includegraphics[width=\textwidth]{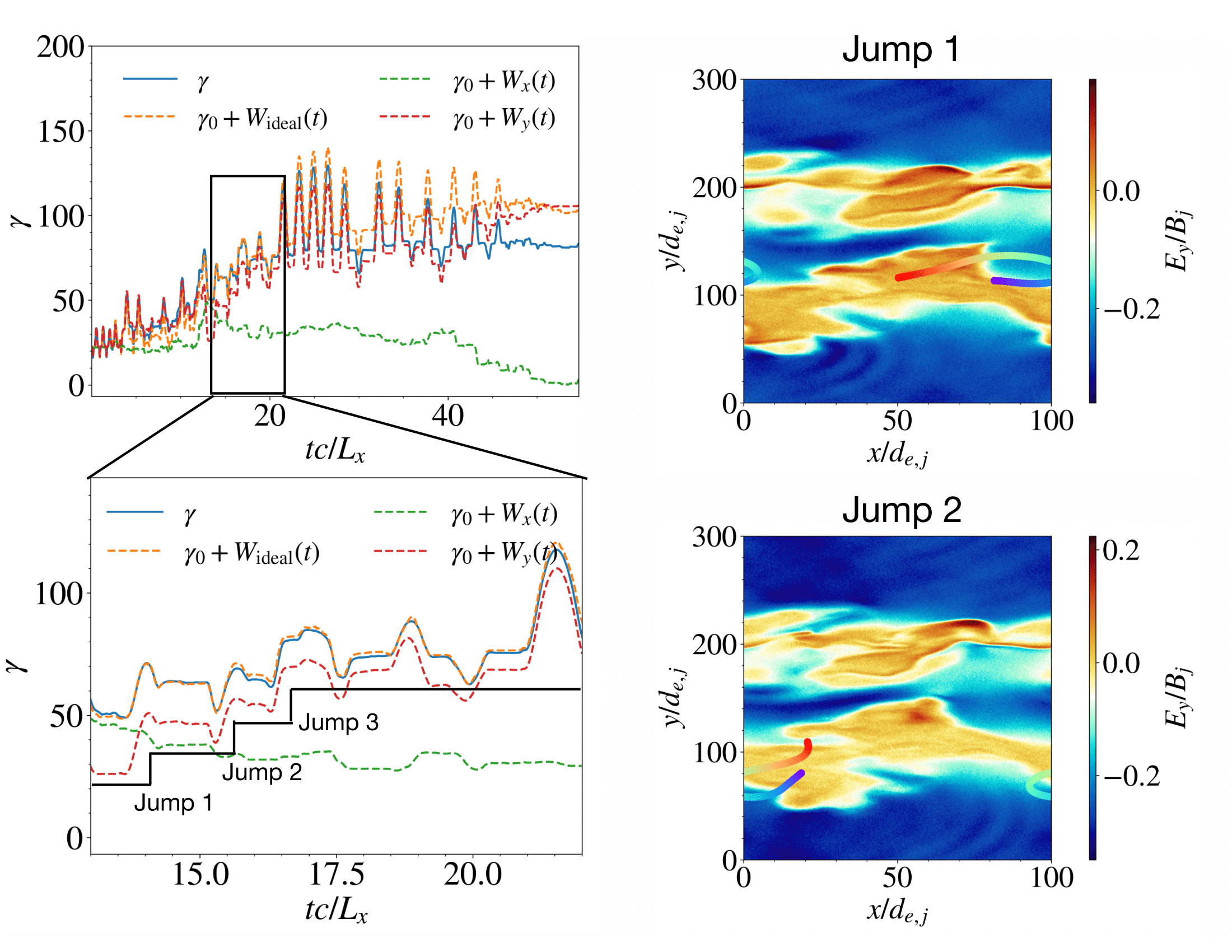}
\caption{Left panels: particle energetics of a selected particle (electron) that has reached a Lorentz factor of $\gamma=95$ at $tc/L_x=27.4$ in the $u_j=0.3$ mixed-shear case, with a zoom-in panel focusing on the time period $13\leq tc/L_x\leq 22$. $W_\mathrm{ideal}, W_x, W_y$ are defined in the main text in \S\ref{subsec:ptcl_acc}. Right panels: Trajectories of the particle over two time windows, corresponding to Jumps 1 and 2 indicated in the lower left panel, overlaid on an $E_y$ snapshot. Colorbar indicates the strength of $E_y$, while the rainbow-colored trajectories indicate the progress of time (blue: early; red: later).}
    \label{fig:uj0.3_jumps_g95}
\end{figure}

\begin{figure}
    \centering
    \includegraphics[width=\textwidth]{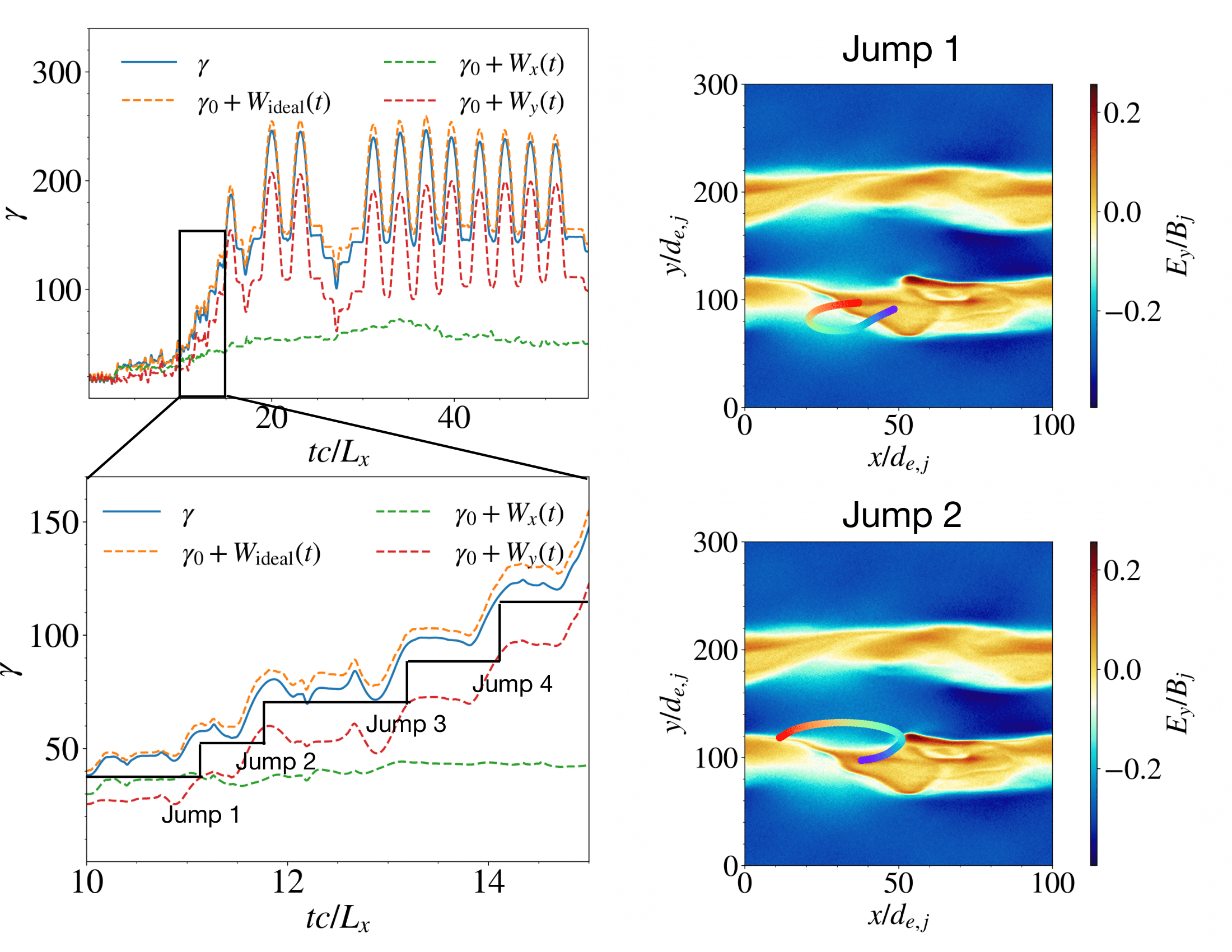}
    \caption{Same as in Fig.~\ref{fig:uj0.3_jumps_g95} but for a particle (electron) that has reached a Lorentz factor of $\gamma=105$ at $tc/L_x=27.4$.}
    \label{fig:uj0.3_jumps_g105}
\end{figure}

In Figs.~\ref{fig:uj0.3_jumps_g95} and \ref{fig:uj0.3_jumps_g105} we show two more examples of particle acceleration through the new mechanism described in \S\ref{subsec:ptcl_acc}, i.e. asymmetric work done by the motional $E_y$ as a particle  performs S-shaped orbits through a bent shear layer, acquiring net energization. In Fig.~\ref{fig:uj0.3_jumps_g95} we select a particle that has reached a Lorentz factor $\gamma=95$ at $tc/L_x=27.4$, while in Fig.~\ref{fig:uj0.3_jumps_g105} we select a particle that has reached a Lorentz factor $\gamma=105$ at $tc/L_x=27.4$. In both figures, we show the S-shaped orbits performed by the particles over two time windows, corresponding to Jumps 1 and 2 indicated in the lower left panel, where the particle acquires net energization from the work done by $E_y$.

\bibliographystyle{jpp}
\bibliography{references}

\end{document}